# Advanced Stationary Point Concentration Technique for Leakage Mitigation and Small Drone Detection with FMCW Radar

Junhyeong Park, *Graduate Student Member*, *IEEE*, Jun-Sung Park,
Kyung-Bin Bae, and Seong-Ook Park, *Senior Member*, *IEEE*

***Abstract*—As the threats of small drones have grown, developing radars to detect the small drones has become an important issue. In earlier studies, we proposed the stationary point concentration (SPC) technique for the small drone detection with frequency-modulated continuous-wave (FMCW) radar. The SPC technique is a new approach to mitigate the leakage that is an inherent problem in the FMCW radar. The SPC technique improves the signal-to-noise ratio of the small drones by reducing the noise floor and provides accurate distance and velocity information of the small drones. However, the SPC technique has shortcomings in realizing it. In this paper, we present the drawbacks of the SPC technique clearly and propose an advanced SPC (A-SPC) technique. The A-SPC technique can overcome the drawbacks of the SPC technique while taking all the good effects of the SPC technique. The experimental results verify the proposed A-SPC technique and show its robustness and usefulness.**

## I. Introduction

TO cope with the growing serious incidents caused by small drones, many radars for detecting the small drones have been developing. Frequency-modulated continuous-wave (FMCW) radar, well-known for its cost-effective and high-resolution remote sensing capability, has been frequently chosen for small drone detection [1]-[9]. However, there is an inherent problem called leakage in the FMCW radar. The leakage signal, which has immense power and poor phase noise, degrades the sensitivity of the FMCW radar by dominating and raising the noise floor [6]-[16].

This work has been submitted to the IEEE for possible publication. Copyright may be transferred without notice, after which this version may no longer be accessible. This work was supported by the Institute for Information and communications Technology Promotion grant funded by the Korea Government (MSIT) under Grant 2018-0-01658, Key Technologies Development for Next Generation Satellites. (*Corresponding author: Junhyeong Park.*)

The authors are with the School of Electrical Engineering, Korea Advanced Institute of Science and Technology, Daejeon 34141, South Korea (e-mail: bdsfh0820@kaist.ac.kr; kirasnip@kaist.ac.kr; carrierbkb@kaist.ac.kr; soparky@kaist.ac.kr).

Many studies have conducted to attenuate the leakage. In [10]-[13], closed-loop leakage cancellers added to the radar system were proposed. Through the closed-loop, an error vector including the amplitude and phase of the leakage is adaptively generated and fed into an RF front-end. In [14]-[16], various balanced topologies introduced in the radar front-end were proposed to cancel out the leakage. The previous techniques were based on the typical approach that creates the same signal as the leakage and subtracts it from the received signal. Also, they required additional hardware parts that are not basic components in FMCW radars.

We proposed the stationary point concentration (SPC) technique, a completely new approach for the leakage mitigation, in earlier studies [6]-[9]. The SPC technique mitigates the leakage by concentrating the phase noise of the leakage on a stationary point of a sinusoidal function. We have proved that the SPC technique not only improves the signal-to-noise ratio (SNR) of the small drones by reducing the noise floor but also corrects the distance and velocity information of them. Additionally, the SPC technique can be realized through digital signal processing (DSP) based on strategic frequency planning and oversampling without additional hardware.

However, there are limitations to the SPC technique. First, strictly speaking, the strategic frequency planning, one of the processes for realizing the SPC technique, limits the freedom in frequency planning when engineers design radar systems. Second, the oversampling may require high-performance analog-to-digital (ADC) converter and memories, which can increase the cost. Finally, the radar architecture to which the SPC technique can be applied is limited. The SPC technique is not available for FMCW radars with homodyne architecture.

In this paper, we clearly describe the limitations of the SPC technique. Then, we propose an advanced SPC (A-SPC) technique to overcome these limitations. The A-SPC technique introduces a quadrature demodulator and complex signal-based DSP to eliminate the cause of limitations in the SPC technique. Moreover, the A-SPC technique includes an algorithm of quadrature imbalance correction to resolve the quadrature imbalance that is a practical problem in quadrature demodulators.

Unlike the SPC technique, the A-SPC technique does not require the strategic frequency planning and the oversampling.

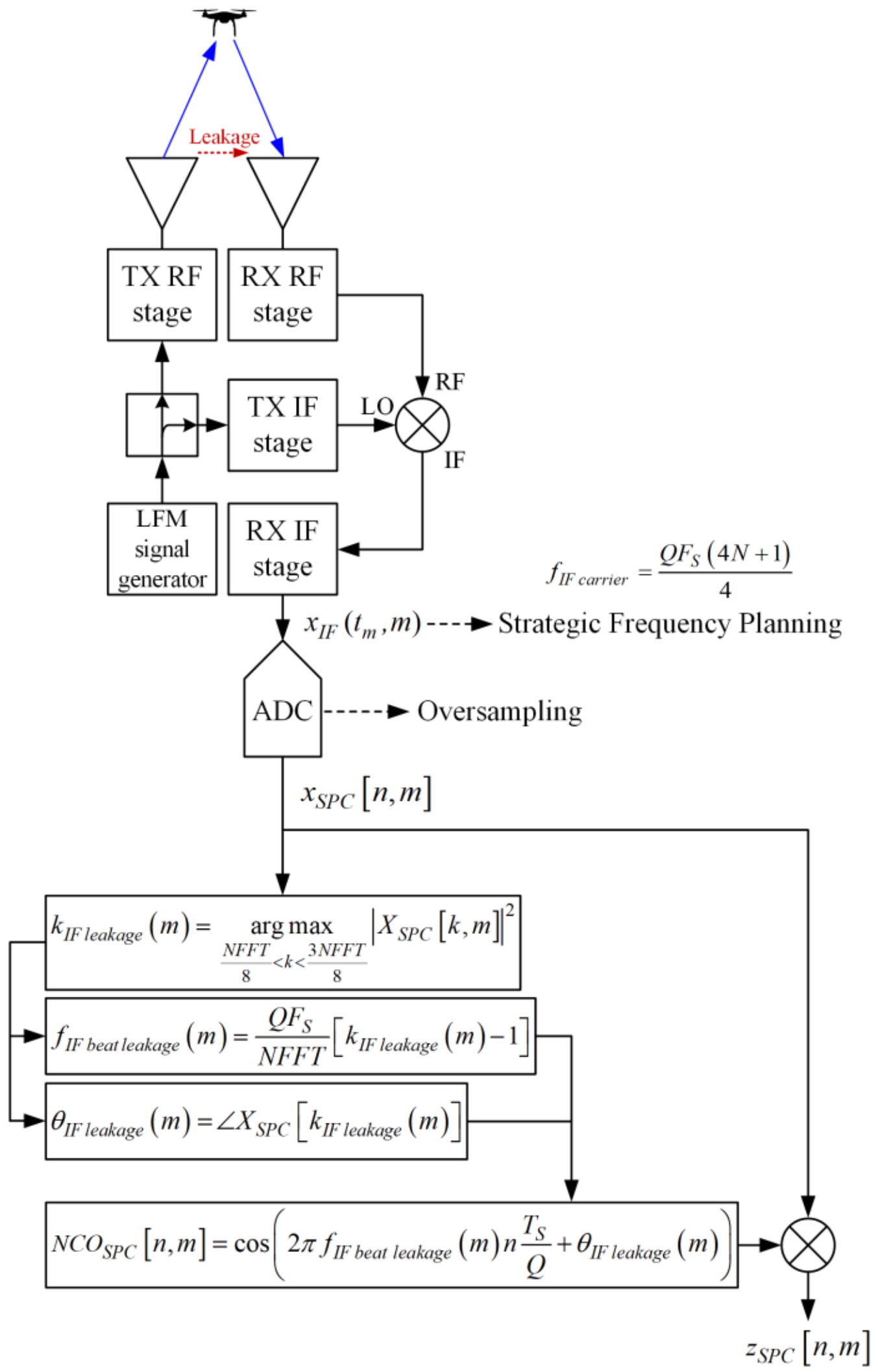

Fig. 1. Block diagram of the SPC technique.

Besides, the A-SPC technique can be applied even for the homodyne FMCW radar. Additionally, all the good effects of the SPC technique, such as the reduction of noise floor and the correction of the distance and velocity of targets, are also valid in the A-SPC technique. In the case of the noise floor reduction, the A-SPC technique can even lower the noise floor more than the SPC technique can do. Also, under the same sampling condition, the maximum unambiguous range (MUR) resulting from the A-SPC technique is wider than twice the MUR resulting from the SPC technique.

We describe theories and realization method of the A-SPC technique in detail. Then, we demonstrate the aforementioned performances of the A-SPC technique with various experiments. For the experiments, both the heterodyne and homodyne FMCW radars were used. DJI Inspire 2 and DJI Spark were used as targeted small drones. All DSPs in this paper were conducted through MATLAB.

In Section II, the theories and realization methods of the SPC technique and the A-SPC technique are explained in detail. In Section III, the experiments and radar systems are introduced. The experimental results are shown and discussed in Section IV. Finally, the conclusion follows in Section V.

## II. Theories and Realization Methods

To clearly present the major points of this paper and reduce the complexity, we skip common mathematical derivation processes before the deramping, such as up-conversions and signal delays in the FMCW radar system. Those processes can also be found in [7], [9].

### A. SPC Technique

Fig. 1 shows a block diagram of the SPC technique. After the deramping process, the IF beat signals, $x_{IF}(t_m, m)$, are extracted and can be written as follows:

$$
\begin{aligned}
&x_{IF}(t_m, m) \\
&= x_{IF\ beat\ leakage}(t_m, m) + x_{IF\ beat\ targets}(t_m, m) \\
&= A_{IF\ leakage} \cos\Bigg( 2\pi(\overbrace{f_{IF\ carrier} + f_{offset} + f_{random,\ ft} + f_{beat\ leakage}}^{f_{IF\ beat\ leakage}(m)})t_m \\
&\qquad + \underbrace{2\pi(f_{IF\ carrier} + f_{offset} + f_{random,\ st})Tm + \theta_{leakage}}_{\theta_{IF\ leakage}(m)} \\
&\qquad + \varphi_{IF\ leakage}(t_m, m) \Bigg) \\
&\quad + \sum_{r=1}^{R} A_{IF\ target,r} \\
&\quad \times \cos\Bigg( 2\pi(\overbrace{f_{IF\ carrier} + f_{offset} + f_{random,\ ft} + f_{beat\ leakage}}^{f_{IF\ beat\ leakage}(m)} + f_{beat\ target,r})t_m \\
&\qquad + 2\pi(f_{IF\ carrier} + f_{offset} + f_{random,st} \pm f_{d,r})Tm + \theta_{target,r} \\
&\qquad + \varphi_{IF\ target,r}(t_m, m) \Bigg), \qquad (1)
\end{aligned}
$$

for $0 < t_m < T$, where $t_m = t - Tm$ and $T$ is the sweep period of linear frequency modulation (LFM) signal. $t_m$ and $m$ represents the fast time domain and the slow time domain, respectively. $A_{IF\ leakage}$ & $A_{IF\ target,r}$ and $\varphi_{IF\ leakage}(t_m, m)$ & $\varphi_{IF\ target,r}(t_m, m)$ are the amplitudes and the phase noises of the leakage and the $r$th target beat signals at the IF stage. $f_{IF\ carrier}$ is the carrier frequency at the IF stage. $f_{offset}$, $f_{random,ft}$, and $f_{random,st}$ are unwanted problematic frequency components that can occur in practical radar systems [9]. $f_{offset}$ is the frequency offset of the carrier frequency. $f_{random,ft}$ and $f_{random,st}$ are randomly changing frequency components in the fast time domain and the slow time domain. $f_{offset}$ and $f_{random,st}$ causes the unwanted Doppler shift, velocity error, in the range-Doppler (r-D) map [9]. $f_{beat\ leakage}$ is the beat frequency of the leakage signal mainly due to internal delays. Because $f_{beat\ leakage}$ is also included in the beat signals of targets, it causes the unwanted range shift, distance error, in the power spectrum and r-D map [7], [9]. $f_{beat\ target,r}$ and $f_{d,r}$ are the beat frequency and Doppler frequency of the $r$th target signal.

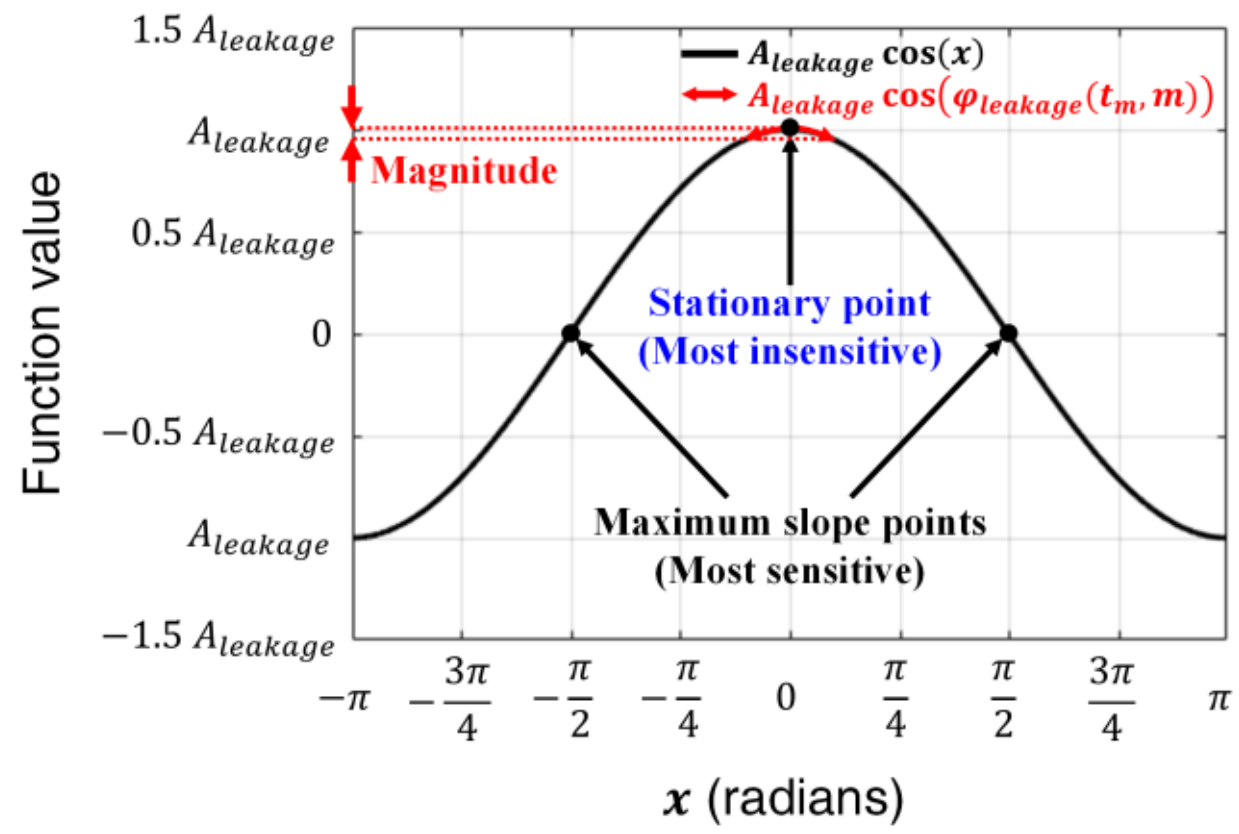

Fig. 2. Key concept of the SPC technique.

As shown in Fig. 2, the key to the SPC technique is to concentrate the phase noise of the leakage at the stationary point of the sinusoidal function. By doing this, the magnitude of the phase noise of the leakage, which manifests as voltage or current noise, is significantly attenuated so that the noise floor decreases [6]-[9]. The SPC technique realizes this effect with a special last down-conversion based on the DSP, as presented in Fig. 1 and Fig. 3. As a result of the multiplication that is the last down-conversion, difference-terms and sum-terms come out. The SPC technique includes the strategic frequency planning and the oversampling to keep the desired domain as far away as possible from the sum-terms and their mirrored signals. Through the strategic frequency planning, $f_{IF\ carrier}$ is placed at a quarter-point of the oversampled frequency domain. If the undersampling is considered additionally, it can be generally said that the strategic frequency planning makes $f_{IF\ carrier}$ placed at $QF_S(4N+1)/4$, where $Q$ is the oversampling factor that is a positive rational number, $N$ is the undersampling factor that is a non-negative integer, and $F_S$ is the minimum available sampling frequency according to the Nyquist theorem. There are various ways of implementing the strategic frequency planning. For example, it can be implemented by adjusting local oscillators (LOs) in the TX RF stage and the RX RF stage without an LO in the RX IF stage. It can also be implemented by additionally placing another LO in the RX IF stage, like the double conversion superheterodyne.

After the oversampling, the oversampled IF beat signals for the SPC technique, $x_{SPC}[n,m]$, can be expressed as follows:

$$
\begin{aligned}
&x_{SPC}[n,m] \\
&= x_{SPC,\ IF\ beat\ leakage}[n,m] + x_{SPC,\ IF\ beat\ targets}[n,m] \\
&= A_{IF\ leakage}\cos\Big(2\pi f_{IF\ beat\ leakage}(m)\, n\frac{T_S}{Q} + \theta_{IF\ leakage}(m) \\
&\qquad + \varphi_{IF\ leakage}\Big(n\frac{T_S}{Q}, m\Big)\Big) \\
&+ \sum_{r=1}^{R} A_{IF\ target,r} \\
&\times\cos\Big(2\pi\big(f_{IF\ beat\ leakage}(m) + f_{beat\ target,r}\big)\, n\frac{T_S}{Q} \\
&\qquad + 2\pi\big(f_{IF\ carrier} + f_{offset} + f_{random,st} \pm f_{d,r}\big)Tm + \theta_{target,r} \\
&\qquad + \varphi_{IF\ target,r}\Big(n\frac{T_S}{Q}, m\Big)\Big),
\end{aligned} \tag{2}
$$

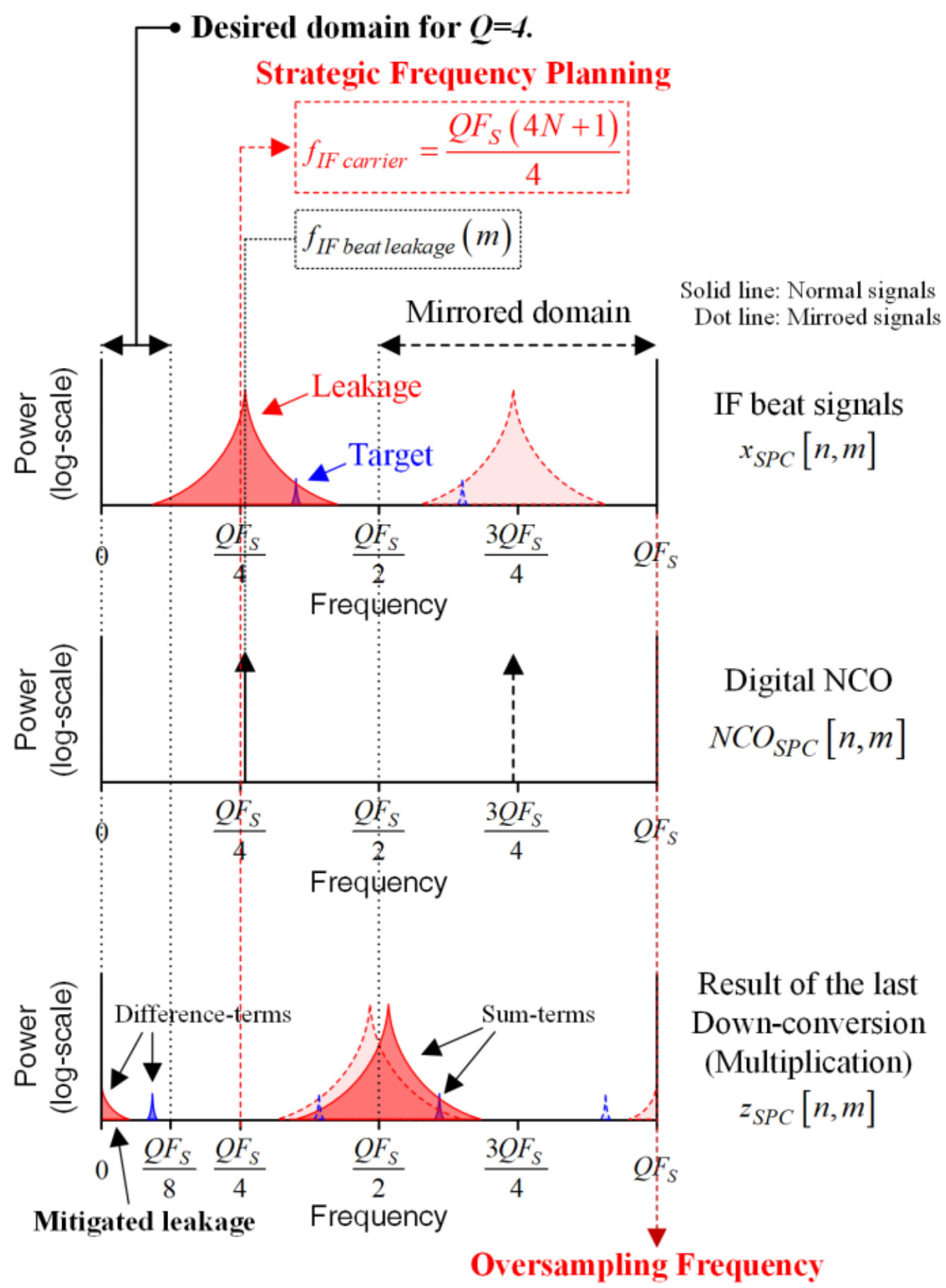

Fig. 3. Conceptual figure for the results of power spectrum for the SPC technique.

where $T_S/Q = 1/QF_S$ is the oversampling interval. For the last down-conversion, the common method naturally uses the LO whose frequency value is $f_{IF\ carrier}$ by considering only the removal of $f_{IF\ carrier}$ that does not have any information. On the other hand, for the last LO, the SPC technique generates a digital numerically controlled oscillator (NCO) whose frequency and phase are the frequency and the phase values of $x_{IF\ beat\ leakage}(t_m, m)$, which are $f_{IF\ beat\ leakage}(m)$ and $\theta_{IF\ leakage}(m)$. These values can be found as follows:

$$
k_{IF\ leakage}(m) = \underset{\frac{NFFT}{8}<k<\frac{3NFFT}{8}}{\arg\max} \left|X_{SPC}[k,m]\right|^2,
$$

$$
f_{IF\ beat\ leakage}(m) = \frac{QF_S}{NFFT}\left[k_{IF\ leakage}(m) - 1\right],
$$

$$\theta_{IF\ leakage}(m) = \angle X_{SPC}\left[k_{IF\ leakage}(m)\right], \tag{3}$$

where $X_{SPC}[k,m]$ is the result of the $NFFT$-point fast Fourier transform (FFT) of $x_{SPC}[n,m]$ along the fast time domain. $NFFT$ is the total number of samples and zero-pads for the zero-padding. $k_{IF\ leakage}(m)$ is the index number that represents the peak for $x_{SPC,IF\ beat\ leakage}[n,m]$, and $\angle X_{SPC}$ is the phase response of $X_{SPC}[k,m]$. Based on found $f_{IF\ beat\ leakage}(m)$ and $\theta_{IF\ leakage}(m)$, the digital NCO is generated as follows:

$$NCO_{SPC}[n,m] = \cos\left(2\pi f_{IF\ beat\ leakage}(m)\, n\frac{T_S}{Q} + \theta_{IF\ leakage}(m)\right), \tag{4}$$

Finally, $x_{SPC}[n,\ m]$ is mixed with the digital NCO, and the desired terms in the final output signals, $z_{SPC}[n,m]$, can be extracted as follows:

$$\begin{aligned} z_{SPC}[n,m] &= z_{SPC,leakage}[n,m] + z_{SPC,targets}[n,m] \\ &= \frac{A_{IF\ leakage}}{2}\cos\left(\varphi_{IF\ leakage}\left(n\frac{T_S}{Q},m\right)\right) \\ &\quad + \sum_{r=1}^{R}\frac{A_{IF\ target,r}}{2} \\ &\quad \times\cos\left(2\pi f_{beat\ target,r}\, n\frac{T_S}{Q} \pm 2\pi f_{d,r} Tm\right. \\ &\qquad \left. + \theta'_{target,r} + \varphi_{IF\ target,r}\left(n\frac{T_S}{Q},m\right)\right), \end{aligned} \tag{5}$$

where $\theta'_{target,r} = \theta_{target,r} - \theta_{leakage}$. Now that the phase noise of the leakage in every $m$th beat signals is concentrated at the stationary point, the noise floors in both the 1-D power spectrum and the 2-D r-D map are significantly reduced so that the SNRs for the targets are improved. In addition, all the unwanted frequency components in the beat signals of targets have gone, thus highly accurate distance and velocity information of the targets can be obtained.

However, strictly speaking, the realization procedures for the SPC technique have limits. Due to the strategic frequency planning, the engineers should design radar systems by considering $f_{IF\ carrier} = QF_S(4N+1)/4$. This causes the limitations not only in the frequency planning but also in the selection of hardware parts. Also, the oversampling itself can require high-performance ADC and memories and cause the increase of the cost. Besides, if the desired digital domain is wide, the required sampling frequency for the oversampling increases and the cost for the ADC and the memories becomes a considerable troublesome. Moreover, since the SPC technique requires the IF stage for the strategic frequency planning, it cannot be applied for the homodyne FMCW radar that does not have the IF stage.

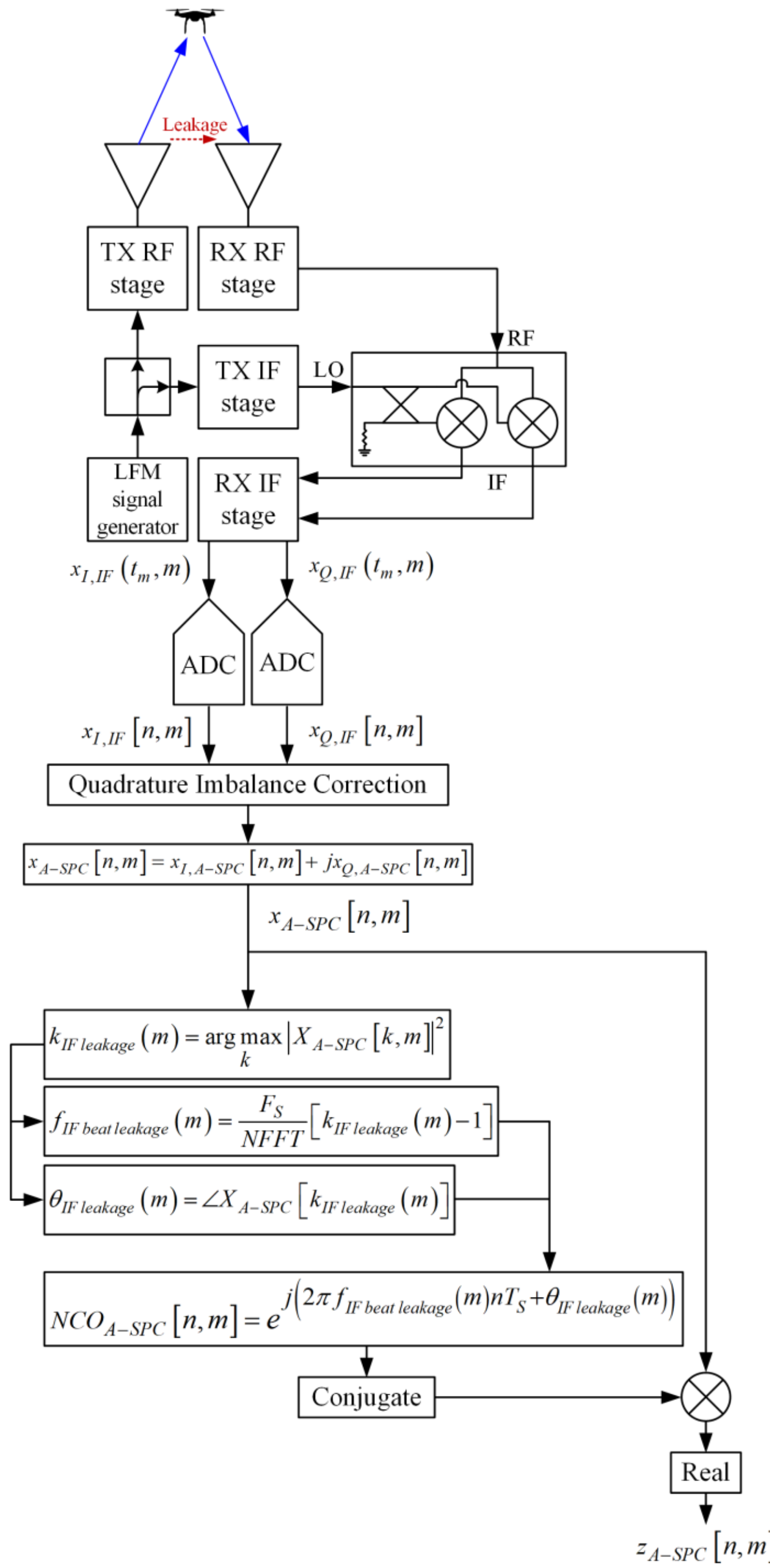

Fig. 4. Block diagram of the A-SPC technique for the heterodyne FMCW radar.

## B. A-SPC Technique

We first describe the case where the A-SPC technique is applied to the heterodyne architecture. Fig. 4 shows the block diagram of the A-SPC technique for the heterodyne FMCW radar. Unlike the SPC technique, a quadrature demodulator is introduced in the A-SPC technique. Also, the DSP is based on complex signals. The quadrature demodulator can also be implemented in the RX IF stage, like the double conversion superheterodyne. Fig. 5 presents how the A-SPC technique works in the heterodyne FMCW radar. Because we use complex-based signals, there are no sum-terms after the last down-conversion. Namely, it is no longer necessary to keep the desired domain as far away as possible from the sum-terms and their mirrored signals. Thus, there is no need to apply the oversampling, and $f_{IF\ carrier}$ can be any value and placed anywhere in the frequency domain, so the engineers can feel free to design radar systems. Therefore, the A-SPC technique can mitigate the leakage without the strategic frequency planning and the oversampling.

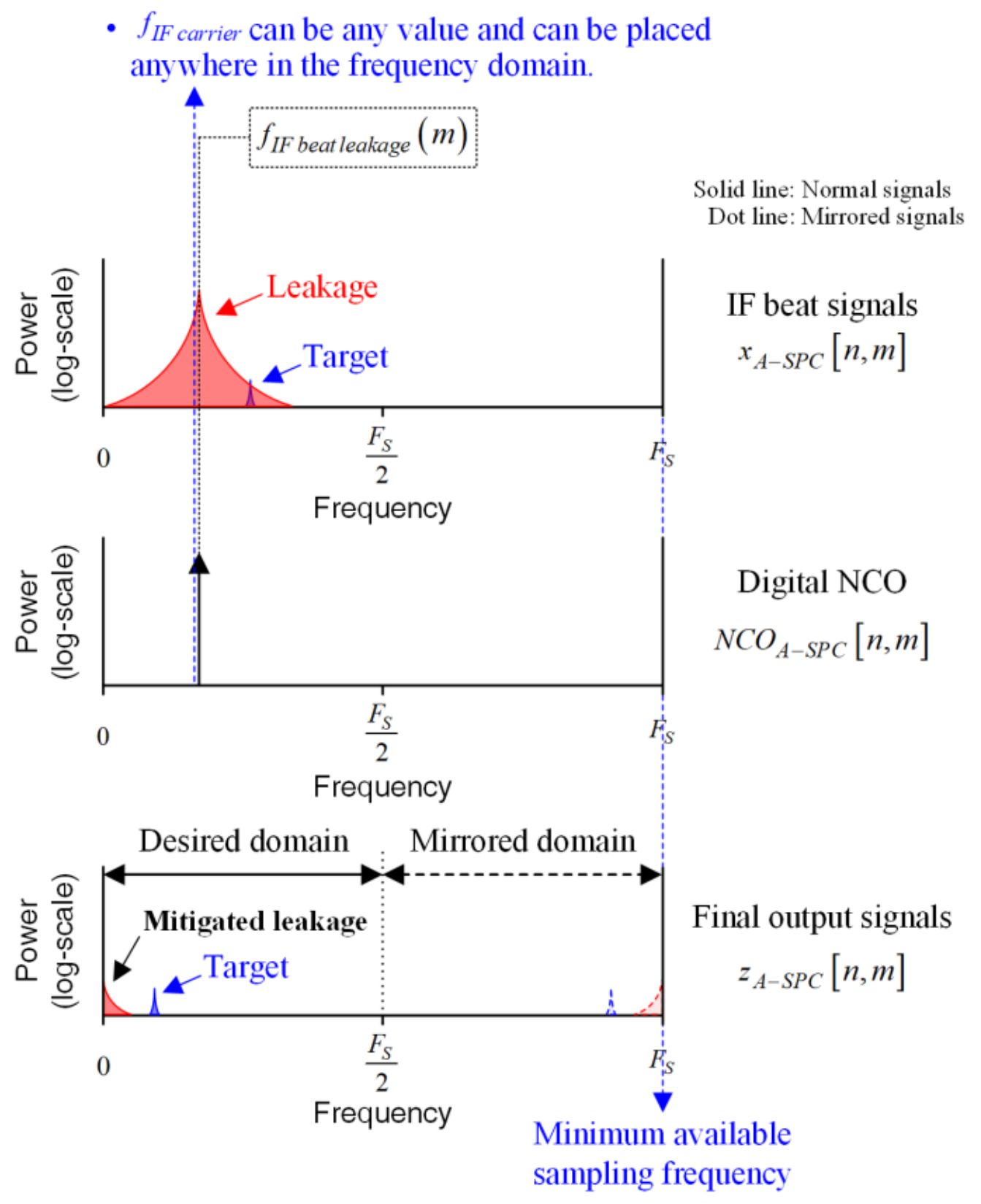

Fig. 5. Conceptual figure for the results of power spectrum for the A-SPC technique when the architecture is heterodyne.

However, the practical quadrature demodulator has a defect called quadrature imbalance. The quadrature imbalance is the amplitude imbalance, $A_E$, and the phase imbalance, $\theta_E$, between the in-phase (I) and quadrature (Q) channels at the IF port. Thus, after the sampling, the I channel, $x_{I,IF}[n,m]$, and the Q channel, $x_{Q,IF}[n,m]$, can be expressed as follows:

$$
\begin{aligned}
x_{I,IF}[n,m] &= A_{IF\ leakage}\cos\Big(2\pi f_{IF\ beat\ leakage}(m)nT_S + \theta_{IF\ leakage}(m) \\
&\qquad + \varphi_{IF\ leakage}(nT_S, m)\Big) \\
&+ \sum_{r=1}^{R} A_{IF\ target,r}\cos\Big(2\pi\big(f_{IF\ beat\ leakage}(m) + f_{beat\ target,r}\big)nT_S \\
&\qquad + 2\pi\big(f_{IF\ carrier} + f_{offset} + f_{random,st} \pm f_{d,r}\big)Tm \\
&\qquad + \theta_{target,r} + \varphi_{IF\ target,r}(nT_S, m)\Big),
\end{aligned}
$$

$$
\begin{aligned}
x_{Q,IF}[n,m] &= A_{IF\ leakage}A_E\sin\Big(2\pi f_{IF\ beat\ leakage}(m)nT_S + \theta_{IF\ leakage}(m) \\
&\qquad + \theta_E + \varphi_{IF\ leakage}(nT_S, m)\Big) \\
&+ \sum_{r=1}^{R} A_{IF\ target,r}A_E\sin\Big(2\pi\big(f_{IF\ beat\ leakage}(m) + f_{beat\ target,r}\big)nT_S \\
&\qquad + 2\pi\big(f_{IF\ carrier} + f_{offset} + f_{random,st} \pm f_{d,r}\big)Tm \\
&\qquad + \theta_E + \theta_{target,r} + \varphi_{IF\ target,r}(nT_S, m)\Big). \qquad (6)
\end{aligned}
$$

The quadrature imbalance induces unwanted image signals, and their magnitude increases as the degree of imbalance increases. Besides, the unwanted image signals can be falsely detected as target signals. To resolve the quadrature imbalance, we included the quadrature imbalance correction as a procedure of the A-SPC technique. Several data-based methods have been proposed for the quadrature imbalance correction [17]-[19]. In [17], algebraic ellipse-fitting and Gram-Schmidt (GS) methods were used for the correction. In [18], geometric ellipse-fitting and GS methods were used for the correction. In [19], only algebraic methods were used to correct the quadrature imbalance. Since these methods require sufficient SNR for the high-quality correction, the authors have used additional hardware such as a metal sphere or metal rod together with a milling machine or motion controller in front of the radar [17]-[19]. However, these external calibrations incur additional costs. Besides, if the imbalance degree is changed, re-calibration with those additional hardware is required.

Unlike the previous papers, an internal calibration method based on the combination of the existing data-based methods is used in this paper. We utilize the leakage signal that has a high power level as a signal that has sufficient SNR for the high-quality correction. In other words, we reverse the disadvantage of the leakage, the inherent problem in the FMCW radar, to an advantage. The quadrature imbalance correction method in this paper applies the geometric ellipse-fitting based on the Taubin algebraic method and the Levenberg-Marquardt algorithm in [18] to $x_{I,IF}[n,m]$ and $x_{Q,IF}[n,m]$ to estimate $A_E$ and $\theta_E$. Then, the quadrature imbalance is corrected by the transform in [19] as follows:

$$
\begin{bmatrix} x_{I,A-SPC}[n,m] \\ x_{Q,A-SPC}[n,m] \end{bmatrix} = \begin{bmatrix} 1 & 0 \\ -\tan(\theta_E) & \dfrac{1}{A_E\cos(\theta_E)} \end{bmatrix} \begin{bmatrix} x_{I,IF}[n,m] \\ x_{Q,IF}[n,m] \end{bmatrix}. \qquad (7)
$$

After the quadrature imbalance correction, the IF beat signals for the A-SPC technique, $x_{A-SPC}[n,m]$, can be written as follows:

$$
\begin{aligned}
x_{A-SPC}[n,m] &= A_{IF\ leakage}e^{j\left(2\pi f_{IF\ beat\ leakage}(m)nT_S + \theta_{IF\ leakage}(m)\right.} \\
&\qquad {}^{\left. + \varphi_{IF\ leakage}(nT_S, m)\right)}
\end{aligned}
$$

$$+\sum_{r=1}^{R} A_{IF\,target,r}$$
$$\times e^{j\left(2\pi\left(f_{IF\,beat\,leakage}(m)+f_{beat\,target,r}\right)nT_S + 2\pi\left(f_{IF\,carrier}+f_{offset}+f_{random,st}\pm f_{d,r}\right)Tm+\theta_{target,r} + \varphi_{IF\,target,r}(nT_S,m)\right)}. \quad (8)$$

The A-SPC technique extracts $f_{IF\ beat\ leakage}(m)$ and $\theta_{IF\ leakage}(m)$ as described in (3) to concentrate the phase noise of the leakage. However, because the frequency planning for $f_{IF\ carrier}$ is no longer limited in the A-SPC technique, the range of peak searching need not be strict. Thus, in the A-SPC technique, $k_{IF\ leakage}(m)$ can be written as follows:

$$k_{IF\,leakage}(m) = \arg\max_k \left|X_{A-SPC}[k,m]\right|^2. \quad (9)$$

After extracting $f_{IF\ beat\ leakage}(m)$ and $\theta_{IF\ leakage}(m)$, the digital NCO for the A-SPC technique is generated as follows:

$$NCO_{A-SPC}[n,m] = e^{j\left(2\pi f_{IF\,beat\,leakage}(m)nT_S+\theta_{IF\,leakage}(m)\right)}. \quad (10)$$

Then, the complex-based last down-conversion is performed by taking the conjugate to $NCO_{A-SPC}[n,m]$ and multiplying it by $x_{A-SPC}[n,m]$. Finally, the final output signals through the A-SPC technique, $z_{A-SPC}[n,m]$, can be extracted as follows:

$$\begin{aligned} z_{A-SPC}[n,m] &= z_{A-SPC,leakage}[n,m] + z_{A-SPC,targets}[n,m] \\ &= A_{IF\,leakage}\cos\left(\varphi_{IF\,leakage}(nT_S,m)\right) \\ &\quad +\sum_{r=1}^{R} A_{IF\,target,r} \\ &\quad \times\cos\left(2\pi f_{beat\,target,r}nT_S \pm 2\pi f_{d,r}Tm + \theta'_{target,r} + \varphi_{IF\,target,r}(nT_S,m)\right). \end{aligned} \quad (11)$$

As shown in (11), the phase noise of the leakage can be concentrated on the stationary point of the sinusoidal function by taking only the real part of the result of the last down-conversion. In addition, all the unwanted frequency components, $f_{offset}$, $f_{random,ft}$, $f_{random,st}$, and $f_{beat\ leakage}$, in the beat signals of targets have removed. Therefore, the A-SPC technique can take all the positive effects of the SPC technique without the strategic frequency planning and the oversampling.

Unlike the SPC technique, because the A-SPC technique has no limitation in determining $f_{IF\ carrier}$, it is also acceptable to make $f_{IF\ carrier}$ zero. Moreover, the A-SPC technique can be applied even in the homodyne FMCW radar that has no IF stage. Fig. 6 shows the block diagram of the A-SPC technique for the homodyne FMCW radar. The quadrature demodulator is included in the RF stage, and received LFM signals are directly converted into the RX BB stage through the deramping. Therefore, the BB beat signals, $x_{I,BB}(t_m,m)$ and $x_{Q,BB}(t_m,m)$, in the homodyne architecture are extracted and can be written as follows:

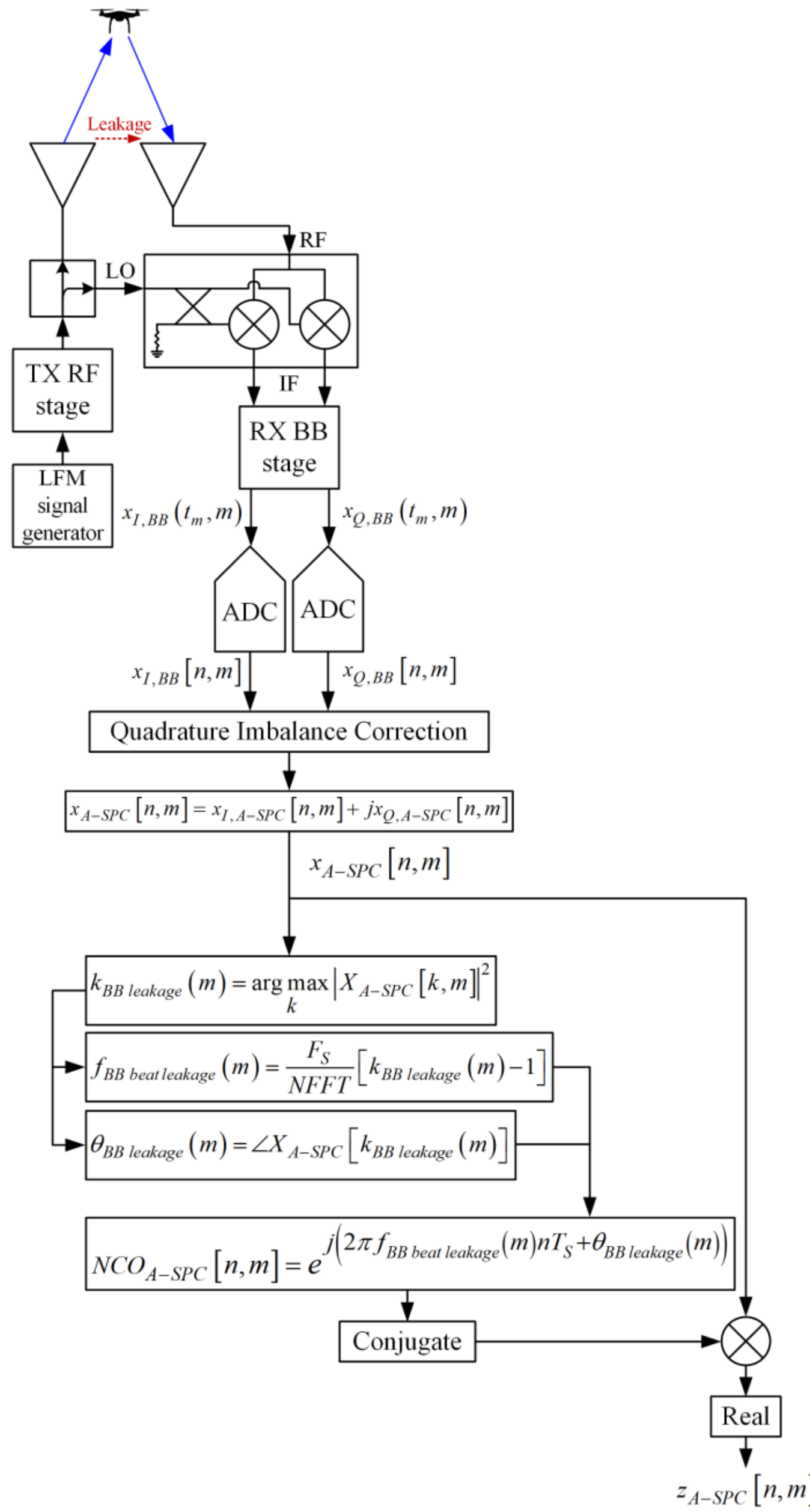

Fig. 6. Block diagram of the A-SPC technique for the homodyne FMCW radar.

$$x_{I,BB}(t_m,m) = A_{BB\,leakage}\cos\Big(2\pi(\overbrace{f_{random,ft}+f_{beat\,leakage}}^{f_{BB\,beat\,leakage}(m)})t_m + \underbrace{2\pi f_{random,st}Tm+\theta_{leakage}}_{\theta_{BB\,leakage}(m)} + \varphi_{BB\,leakage}(t_m,m)\Big)$$

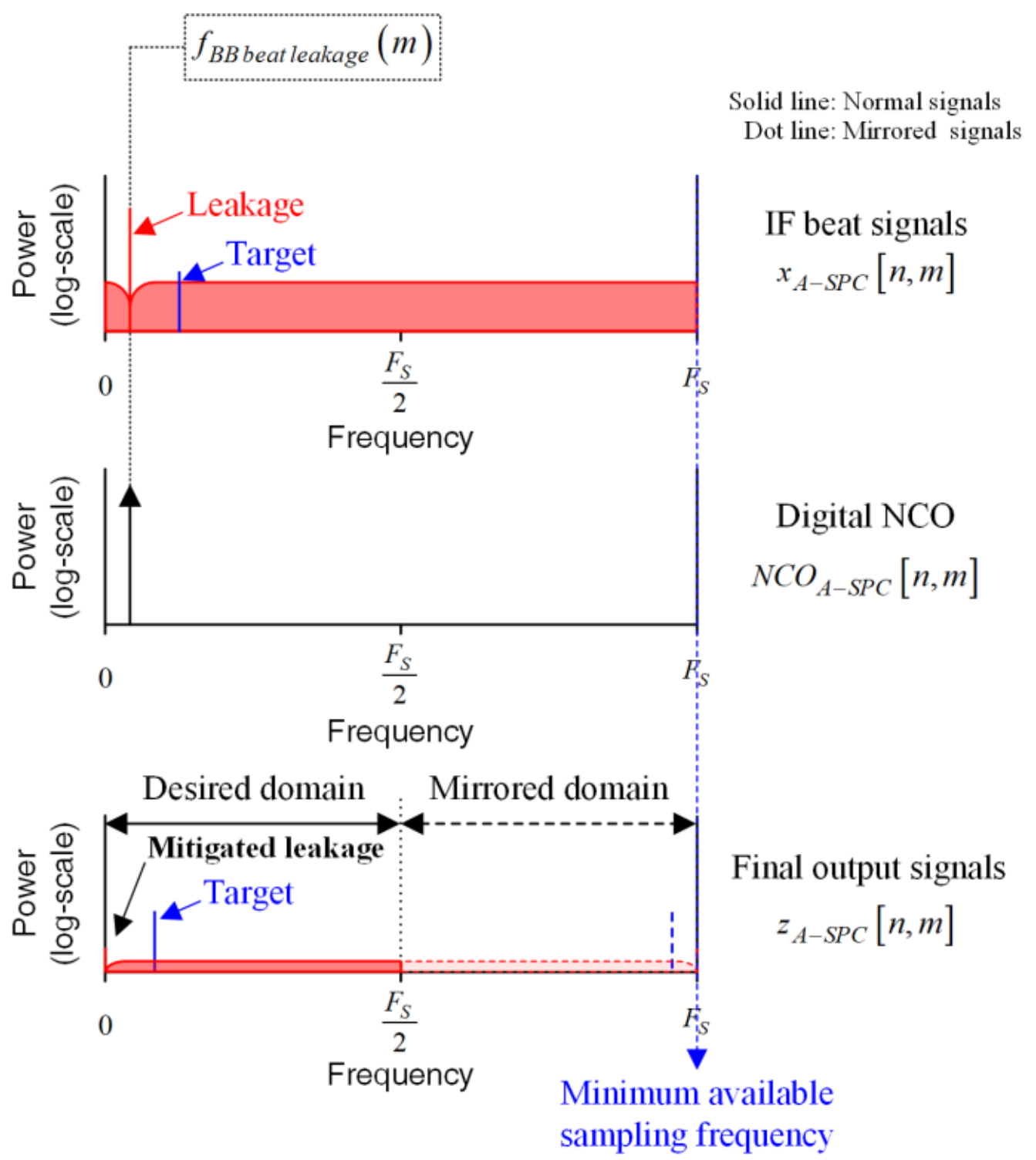

Fig. 7. Conceptual figure for the results of power spectrum for the A-SPC technique when the architecture is homodyne.

$$
\begin{aligned}
&+\sum_{r=1}^{R} A_{BB\,target,r} \\
&\times\cos\Bigg( 2\pi(\overbrace{f_{random,\,ft} + f_{beat\,leakage}}^{f_{BB\,beat\,leakage}(m)} + f_{beat\,target,r})t_m \\
&\qquad +2\pi(f_{random,st} \pm f_{d,r})Tm + \theta_{target,r} + \varphi_{BB\,target,r}(t_m,m) \Bigg),
\end{aligned}
$$

$$
\begin{aligned}
&x_{Q,BB}(t_m,m) \\
&= A_{BBleakage}A_E \sin\Big(2\pi f_{BB\,beat\,leakage}(m)t_m \\
&\qquad +\theta_{BB\,leakage}(m)+\theta_E+\varphi_{BB\,leakage}(t_m,m)\Big) \\
&+\sum_{r=1}^{R} A_{BB\,target,r}A_E \\
&\times\sin\Big(2\pi\big(f_{BB\,beat\,leakage}(m)+f_{beat\,target,r}(m)\big)t_m \\
&\quad +2\pi\big(f_{random,st} \pm f_{d,r}\big)Tm+\theta_{target,r} \\
&\quad +\theta_E+\varphi_{BB\,target,r}(t_m,m)\Big),
\end{aligned} \tag{12}
$$

where $A_{BB\,leakage}$ & $A_{BB\,target,r}$ and $\varphi_{BB\,leakage}(t_m,m)$ & $\varphi_{BB\,target,r}(t_m,m)$ are the amplitudes and the phase noises of the leakage and the $r$th target beat signals at the BB stage in the homodyne architecture. Because the homodyne architecture performs the direct down-conversion in the deramping process, $f_{IF\,carrier}$ and $f_{offset}$ are removed naturally. Also, for the same reason, $\varphi_{BB\,leakage}(t_m,m)$ and $\varphi_{BB\,target,r}(t_m,m)$ have the range correlation effect (RCE) [20]. Thus, the shape of the phase noise can be changed from the skirt shape, and an example shape for this effect is reflected in Fig. 7. Fig. 7 shows how the A-SPC technique works in the homodyne FMCW radar. The rest procedures and principles of the A-SPC technique for the homodyne architecture are the same as those for the heterodyne architecture. In this way, the A-SPC technique can be applied well even for the homodyne FMCW radar, thus it can provide significant improvement in the SNR and the accuracy of the measured distance and velocity of the targets.

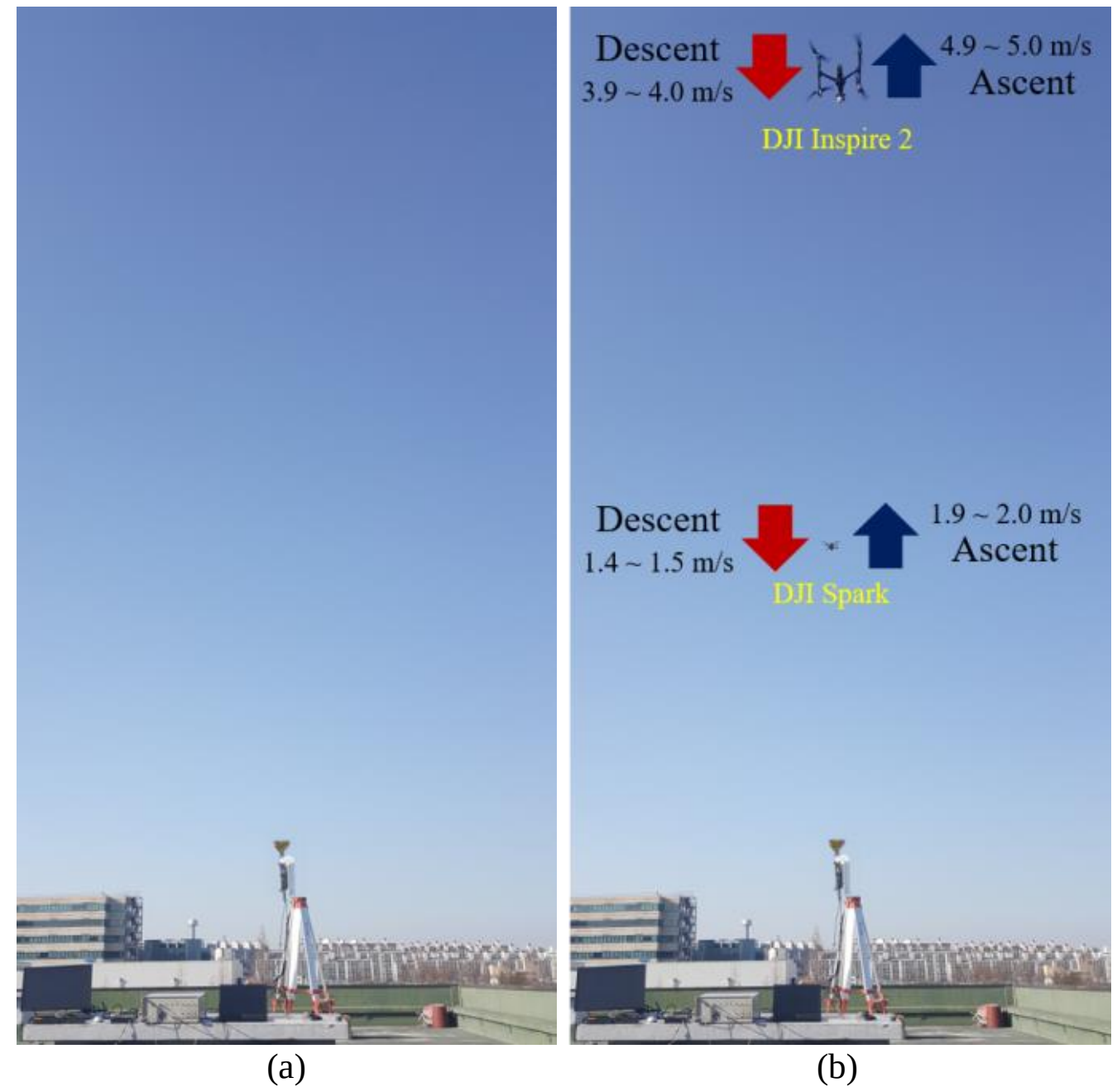

Fig. 8. Experimental scenes of *Experiment A* and *Experiment B.* (a) Radar was operated without the small drones in *Experiment A* and *Experiment B.* (b) Radar was operated with the small moving drones in *Experiment B*.

## III. Experiments and Radar Systems

To verify the A-SPC technique, we carried out three experiments by using both the heterodyne and homodyne FMCW radars. *Experiment A* and *Experiment B* verify the A-SPC technique in the heterodyne FMCW radar, and *Experiment C* validates the A-SPC technique in the homodyne FMCW radar. All the experiments were conducted on the rooftop of a building. For the small drones, we used DJI Inspire 2 and DJI Spark. When moving the drones, we raised and lowered DJI Inspire 2 at the speed of 4.9-5.0 m/s and 3.9-4.0 m/s, respectively. In the case of DJI Spark, we raised and lowered it at the speed of 1.9-2.0 m/s and 1.4-1.5 m/s, respectively. We checked the speed information through a reliable smart mobile application provided by DJI to control drones. For a clear comparison of the power spectrum, we took an average. Each resulting power spectrum is the average of 100 power spectra. Details of each experiment and each radar system are given below.

### A. Experiment A

In *Experiment A*, we proved that the A-SPC could be realized in the heterodyne FMCW radar without the strategic frequency planning and the oversampling. We compared the three power spectra. One is the power spectrum with no technique applied, another is the power spectrum through the SPC technique, and the other is the power spectrum through the A-SPC technique. Mainly, the noise floor and the degree of improvement were observed and compared. For clear observation, we operated the radar without the small drones and received only the leakage signal. Fig. 8(a) shows the experimental scene for *Experiment A*. The *Ku*-band heterodyne FMCW radar in [7] was used. The specifications of the radar are listed in Table I. Also, the differences of the parameter values between the SPC technique and the A-SPC technique for *Experiment A* can be checked in Table I.

TABLE I
SPECIFICATIONS AND PARAMETERS OF FMCW RADAR FOR *EXPERIMENT A*

| Parameters | SPC | A-SPC |
|---|---|---|
| Radar system architecture | Heterodyne | |
| Radar configuration | Quasi-monostatic | |
| Operating frequency | 14.35-14.50 GHz | |
| Transmit power | 30 dBm | |
| Antenna gain | 16 dBi | |
| Sweep bandwidth ($BW$) | 150 MHz | |
| True range resolution | 1 m | |
| Sweep period ($T$) | 880 us | |
| Desired digital bandwidth | 1.25 MHz | |
| Minimum available sampling frequency ($F_S$) | 2.5 MHz | |
| Oversampling factor ($Q$) | 4 | - |
| Oversampling frequency ($QF_S$) | 10 MHz | - |
| Undersampling factor ($N$) | 0 | - |
| # of samples in a chirp | 8800 | 2200 |
| # of samples after discarding early part in a chirp | 8192 | 2048 |
| IF carrier frequency ($f_{IF\ carrier}$) | 2.5 MHz | 0 MHz |
| Window | Hann | |
| $NFFT$ for finding out $f_{IF\ beat\ leakage}$ and $\theta_{IF\ leakage}$ | $2^{20}$ | $2^{18}$ |
| Desired maximum detectable range | 1100 m | |
| Apparent range resolution | 1.074 m | |

TABLE II
SPECIFICATIONS AND PARAMETERS OF FMCW RADAR FOR *EXPERIMENT B*

| Parameters | SPC | A-SPC |
|---|---|---|
| Radar system architecture | Heterodyne | |
| Radar configuration | Quasi-monostatic | |
| Operating frequency | 14.35-14.50 GHz | |
| Transmit power | 30 dBm | |
| Antenna gain | 16 dBi | |
| Sweep bandwidth ($BW$) | 150 MHz | |
| True range resolution | 1 m | |
| Sweep period ($T$) | 880 us | |
| Desired digital bandwidth | 1.25 MHz | |
| Minimum available sampling frequency ($F_S$) | 2.5 MHz | |
| Oversampling factor ($Q$) | 4 | |
| Oversampling frequency ($QF_S$) | 10 MHz | |
| Undersampling factor ($N$) | 0 | |
| # of samples in a chirp | 8800 | |
| # of samples after discarding early part in a chirp | 8192 | |
| IF carrier frequency ($f_{IF\ carrier}$) | 2.5 MHz | |
| Window | Hann | |
| $NFFT$ for finding out $f_{IF\ beat\ leakage}$ and $\theta_{IF\ leakage}$ | $2^{20}$ | |
| Desired maximum detectable range | 1100 m | |
| Apparent range resolution | 1.074 m | |
| # of chirps in a r-D map | 256 | |
| Velocity resolution | 0.0462 m/s | |

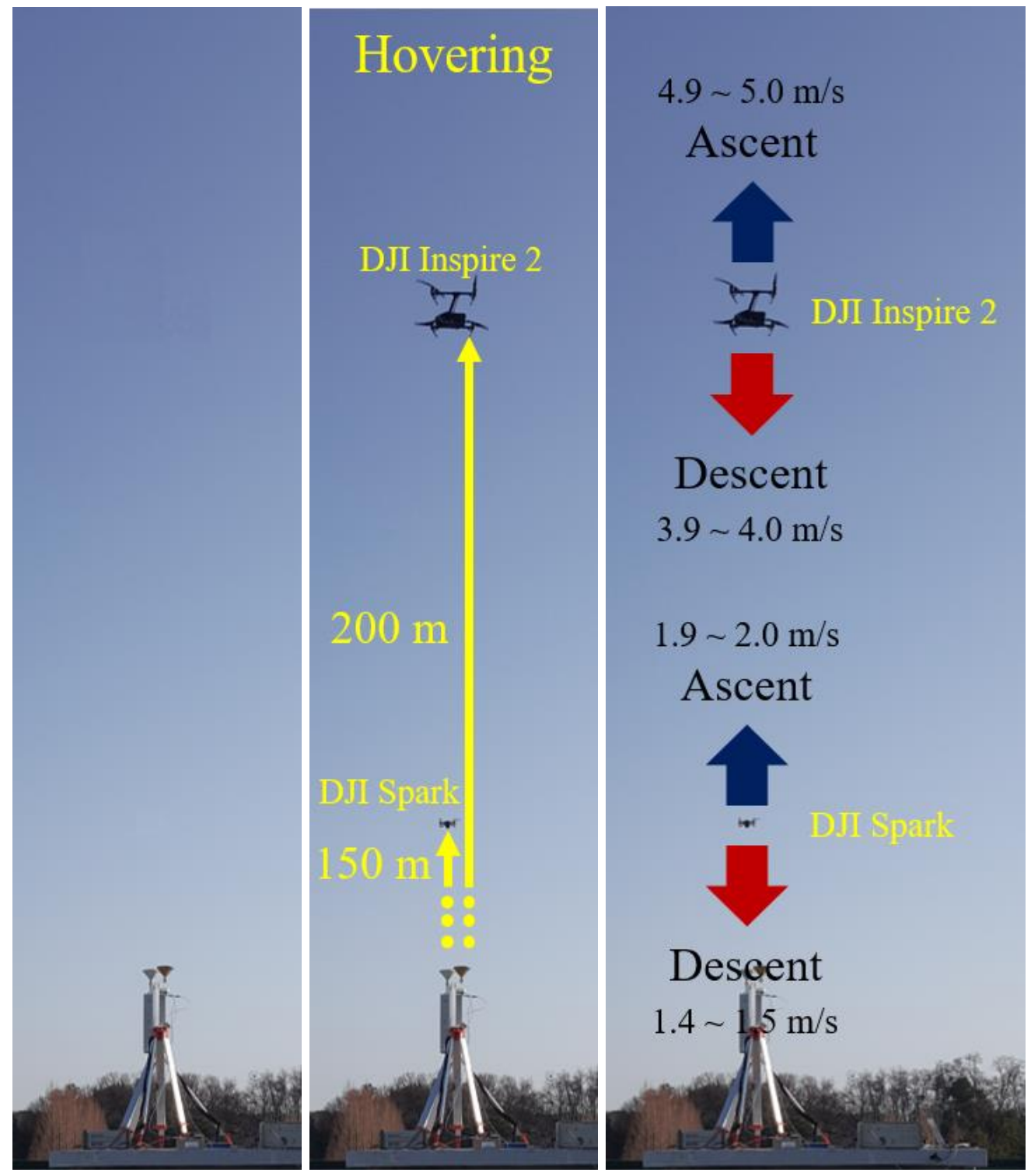

(a) (b) (c)

Fig. 9. Experimental scenes of *Experiment C*. (a) Radar was operated without the small drones. (b) Radar was operated with the small hovering drones. (c) Radar was operated with the small moving drones.

## B. *Experiment B*

In *Experiment B*, we confirmed the effects of the A-SPC technique if it was applied under the same conditions as the SPC technique. In other words, although the A-SPC technique does not require the strategic frequency planning and the oversampling, we checked what effects can be achieved if these are applied. For *Experiment B*, at first, we received only the leakage signal for the clear observation as like Fig. 8(a). We extracted both the power spectra and range-Doppler (r-D) maps through the SPC technique and the A-SPC technique, then we compared them.

Second, as shown in Fig. 8(b), we operated the radar with the small moving drones, and extracted r-D maps. Then, we verified whether the A-SPC technique improves the r-D map by increasing the 2-D SNR and the accuracy of the velocity information of the small moving drones. However, since the *Ku*-band heterodyne FMCW radar has little unwanted problematic frequency components, it is difficult to check whether the A-SPC technique corrects the unwanted Doppler shift. Therefore, we replaced a reference oscillator for one LO inside the radar system with another reference oscillator that is different from the shared reference oscillator. By doing this, only the unwanted Doppler shift occurs in the heterodyne FMCW radar while all the other things remain same. The specifications and the parameters of the radar for *Experiment B* are listed in Table II.

## C. *Experiment C*

In *Experiment C*, we verify whether the A-SPC technique can be applied even in the homodyne FMCW radar. Fig. 9 shows the experimental scenes of *Experiment C*. First, we received only the leakage signal to clearly see the noise floor. Then, we checked that the phase noise of the leakage to which the RCE is applied dominates the noise floor. Also, we confirmed whether the A-SPC technique reduces the noise floor in the power spectrum by mitigating the leakage even if its phase noise shape changes. Second, by placing small drones at certain heights, we demonstrated whether the A-SPC technique improves the SNR and the accuracy of the distance information for the small drones. Finally, we verified whether the A-SPC technique improves the r-D map by increasing the 2-D SNR of the small moving drones. A *Ku*-band homodyne FMCW radar was used for *Experiment C*. Like the *Ku*-band heterodyne FMCW radar used for *Experiment B*, the *Ku*-band homodyne FMCW radar has little unwanted problematic frequency components. Fig. 10 shows its block diagram. In the radar system, Analog Devices AD9854 was included as a direct digital synthesizer (DDS) to achieve high linearity in generating LFM signal. Analog Devices HMC8191 was used for the quadrature demodulator. For the data acquisition (DAQ), Pico Technology PicoScope 4424 was used. The specifications and the parameters of the radar are listed in Table III.

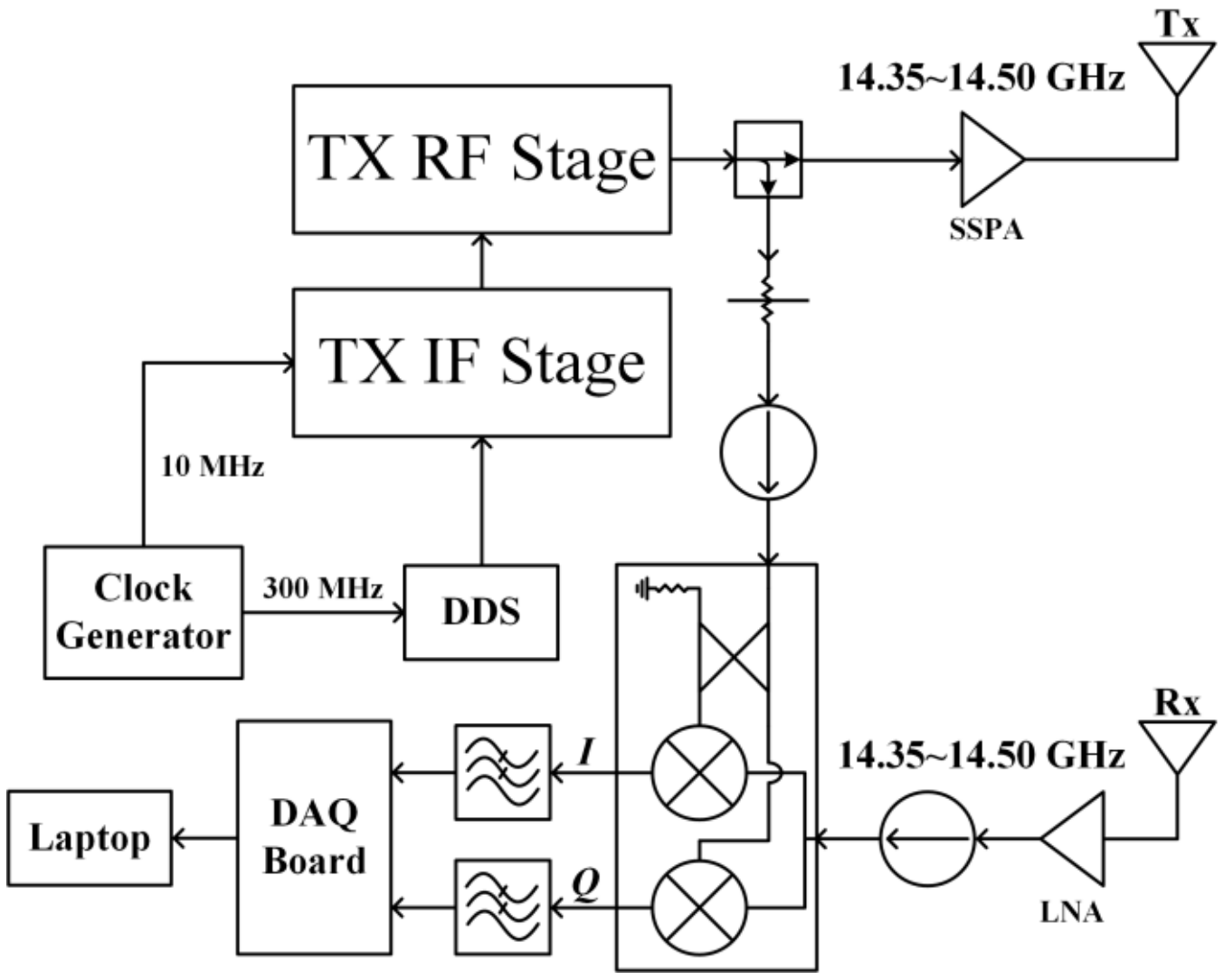

Fig. 10. Block diagram of the *Ku*-band homodyne FMCW radar system.

TABLE III
SPECIFICATIONS AND PARAMETERS OF FMCW RADAR FOR *EXPERIMENT C*

| Parameters | A-SPC |
|---|---|
| Radar system architecture | Homodyne |
| Radar configuration | Quasi-monostatic |
| Operating frequency | 14.35-14.50 GHz |
| Transmit power | 42 dBm |
| Antenna gain | 16 dBi |
| Sweep bandwidth ($BW$) | 150 MHz |
| True range resolution | 1 m |
| Sweep period ($T$) | 500 us |
| Desired digital bandwidth | 2.5 MHz |
| Minimum available sampling frequency ($F_S$) | 5 MHz |
| Oversampling factor ($Q$) | - |
| Oversampling frequency ($QF_S$) | - |
| Undersampling factor ($N$) | - |
| # of samples in a chirp | 2500 |
| # of samples after discarding early part in a chirp | 2048 |
| IF carrier frequency ($f_{IF\ carrier}$) | - |
| Window | Hann |
| $NFFT$ for finding out $f_{IF\ beat\ leakage}$ and $\theta_{IF\ leakage}$ | $2^{19}$ |
| Desired maximum detectable range | 1250 m |
| Apparent range resolution | 1.221 m |
| # of chirps in a r-D map | 256 |
| Velocity resolution | 0.0812 m/s |

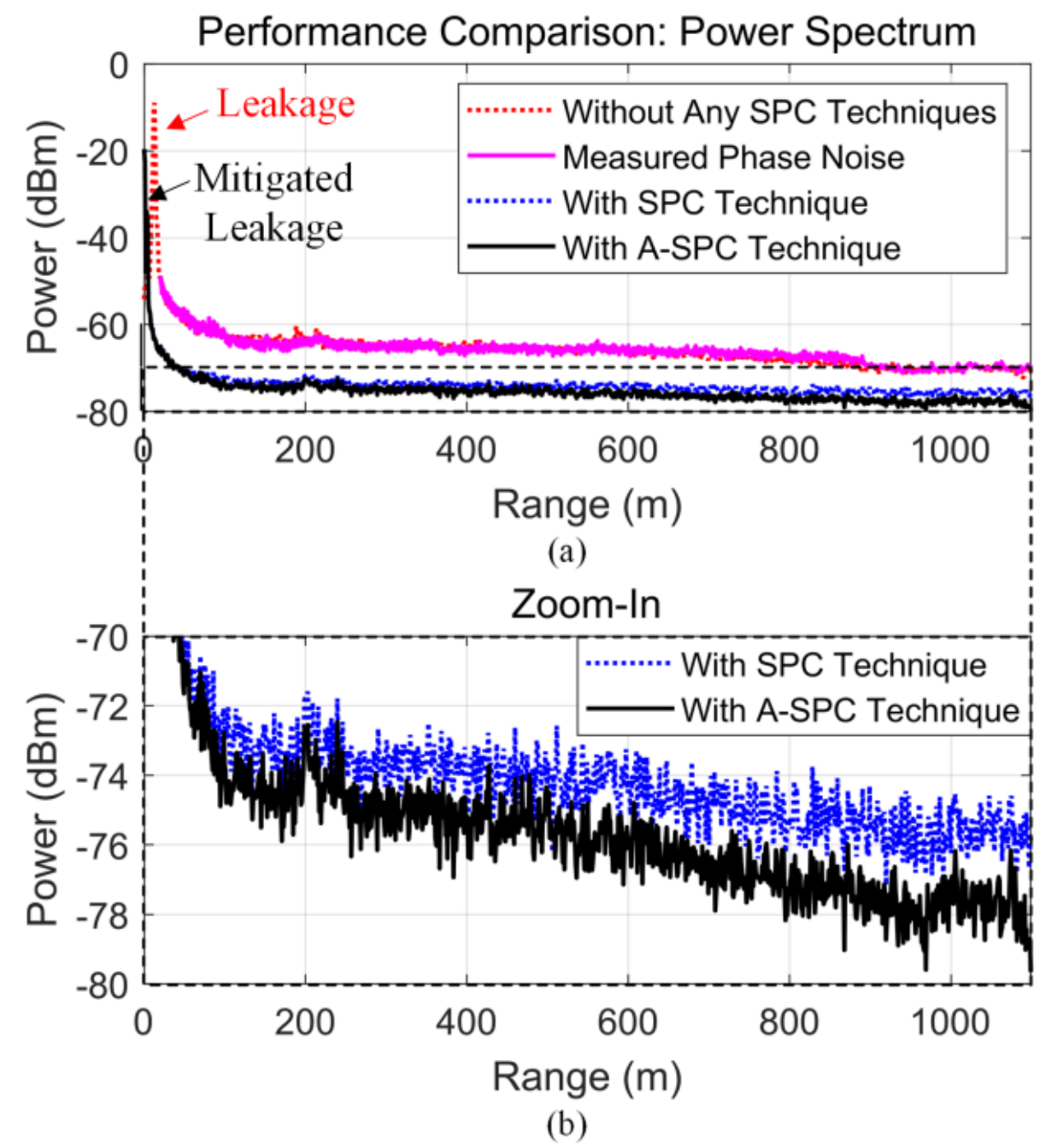

Fig. 11. Results of *Experiment A*: Power spectra. (a) Performance comparison of the power spectra. (b) Zoomed in version of (a).

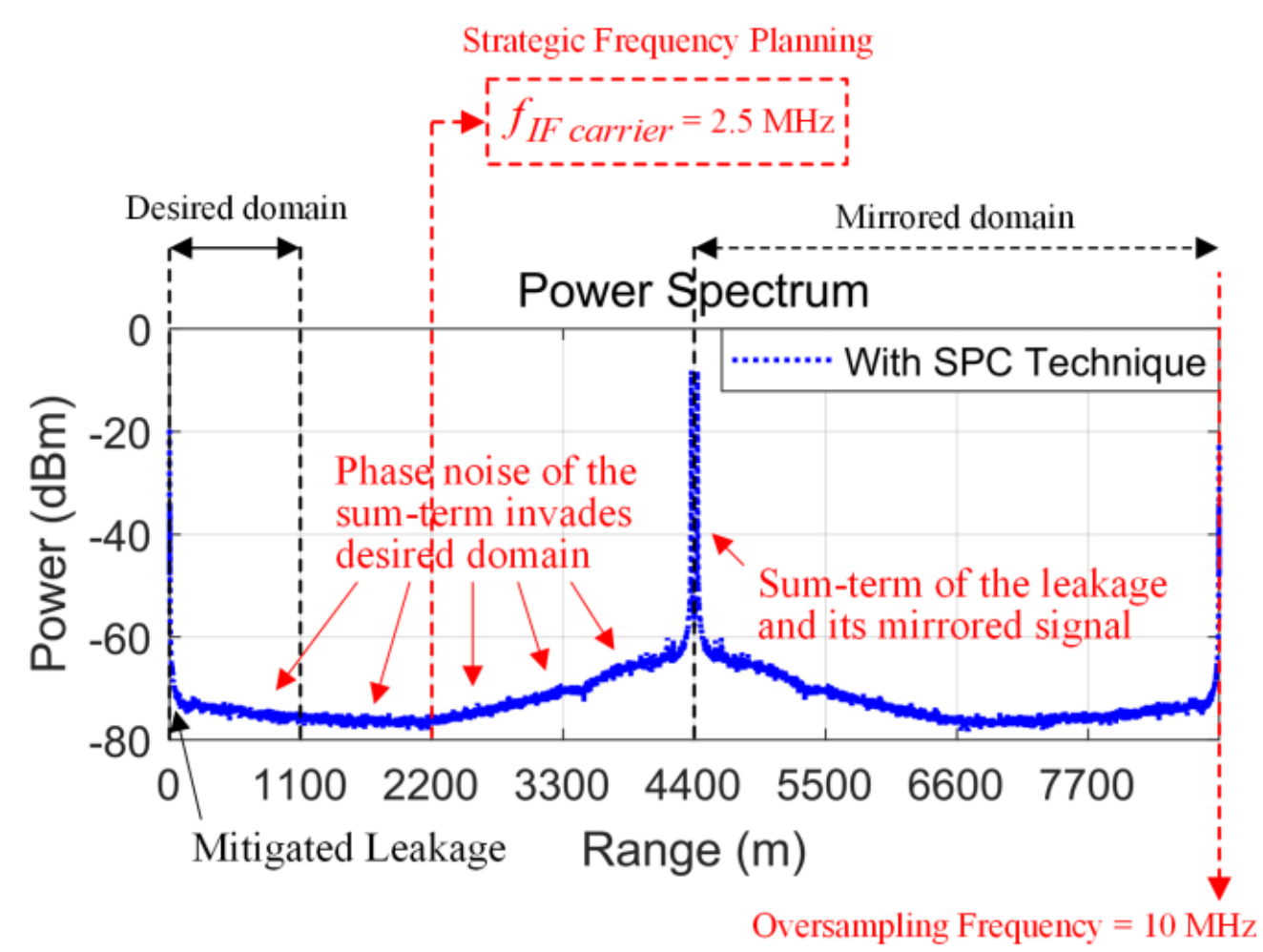

Fig. 12. Results of *Experiment A*: Power spectrum through the SPC technique. The entire domain of the power spectrum is shown.

## IV. RESULTS AND DISCUSSION

### A. *Results and Discussion for Experiment A*

Fig. 11(a) shows the resulting power spectra for *Experiment A*. Because the noise floor is well matched with the measured phase noise, it is confirmed that the phase noise of the leakage dominates the noise floor. Despite the absence of the strategic frequency planning and the oversampling, the A-SPC technique mitigates the leakage and significantly reduces the noise floor. Moreover, the A-SPC technique lowers the noise floor more than the SPC technique does, as shown in Fig. 11(b). This can be explained through Fig. 12. Although the SPC technique puts the sum-term and its mirrored signal far enough away from the desired domain, the phase noise of the sum-term invades the desired domain. Therefore, the noise floor of the power spectrum through the A-SPC technique is lower than that of the power spectrum through the SPC technique.

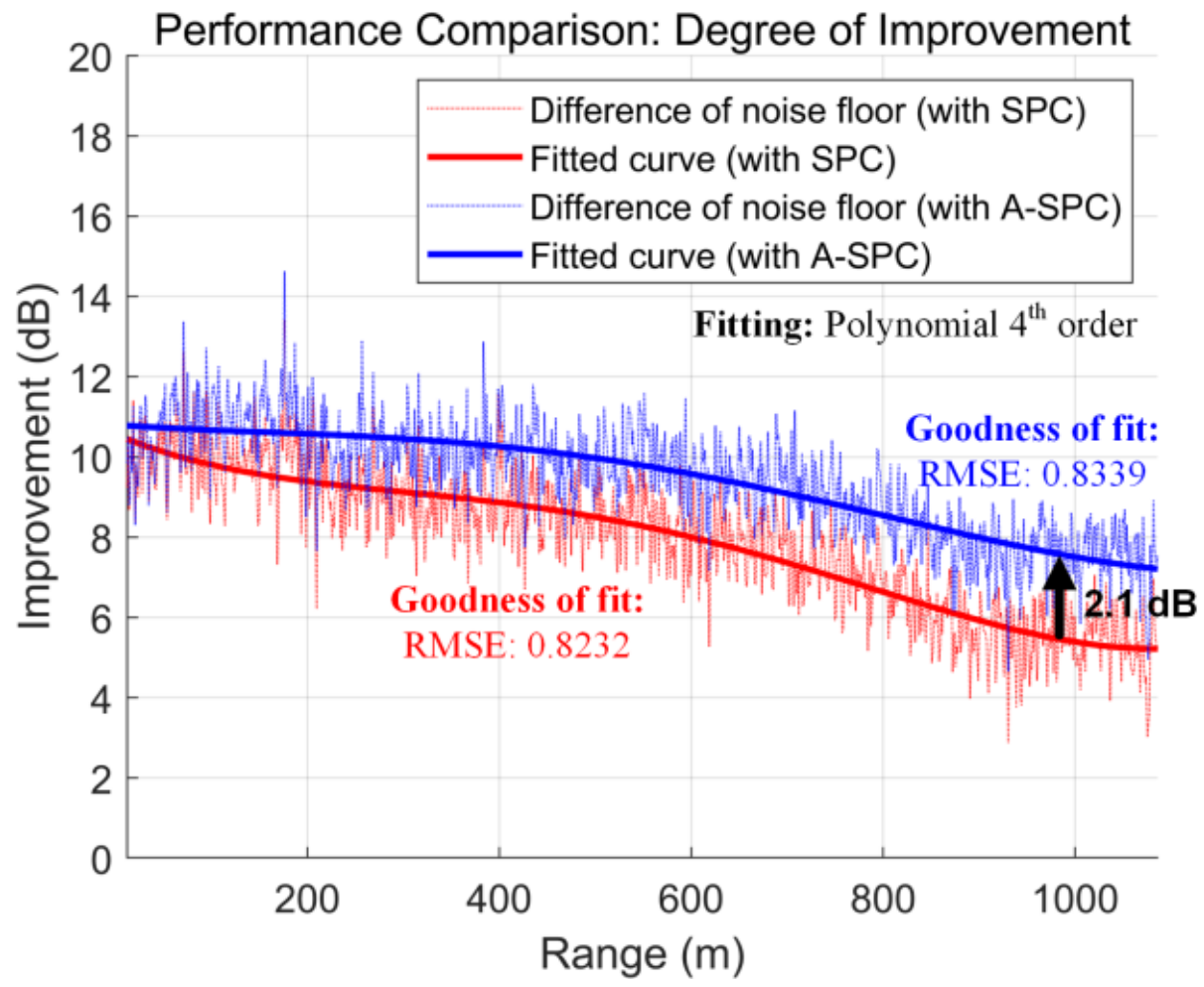

Fig. 13. Results of *Experiment A*: Degrees of improvement for the SPC and the A-SPC technique.

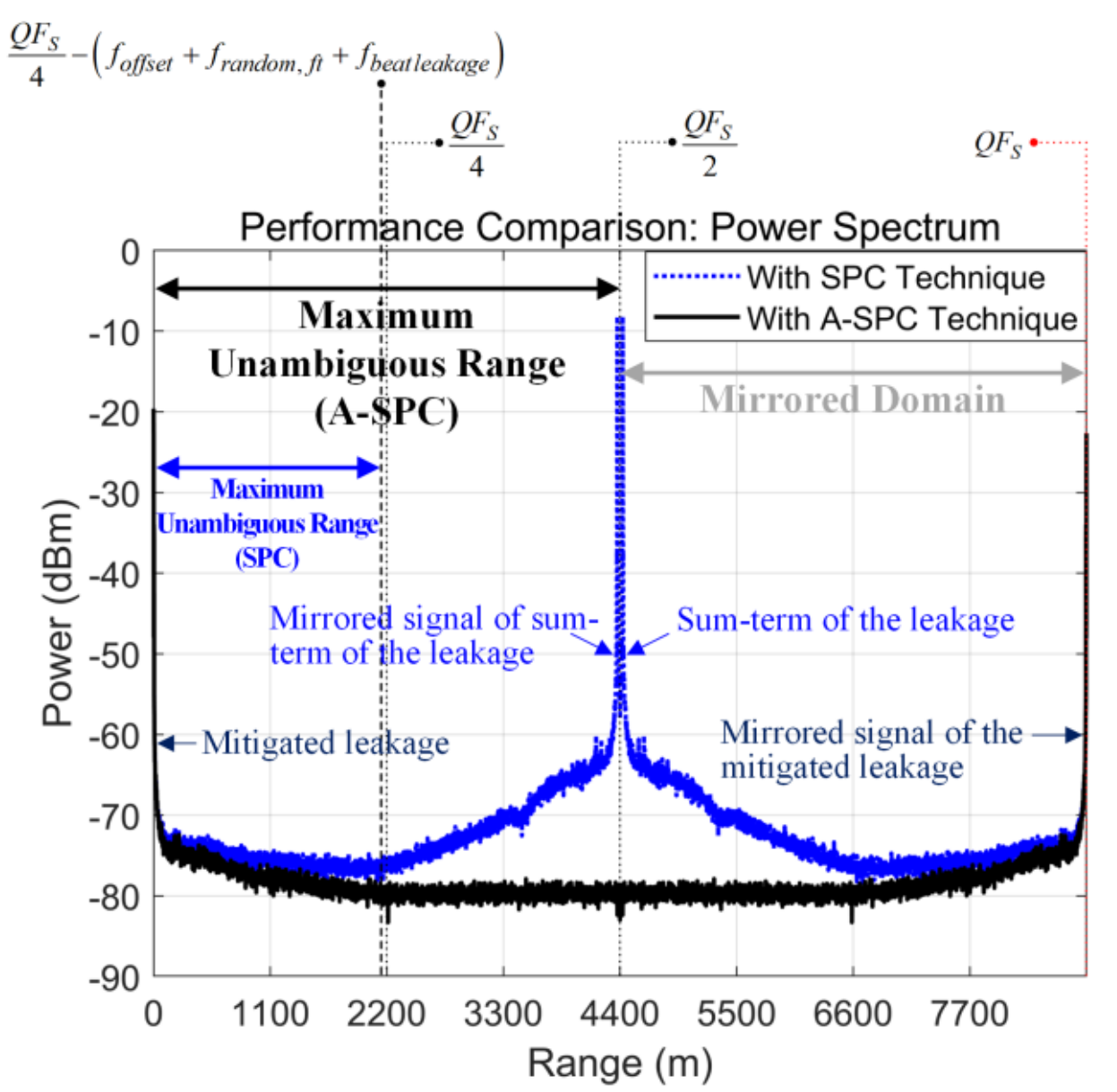

Fig. 14. Results of *Experiment B*: Power spectra. The entire domain of the power spectra is shown.

The degree of improvement for each technique is shown in Fig. 13. After obtaining the differences of the noise floor by subtracting power spectra, we overlaid fitted curves. The degree of improvement through the A-SPC technique is better than that through the SPC technique over the entire desired domain. The increase in the degree of improvement gets better as it goes into the long-range area. In this experiment, the maximum increase in the degree of improvement recorded about 2.1 dB in the long-range area.

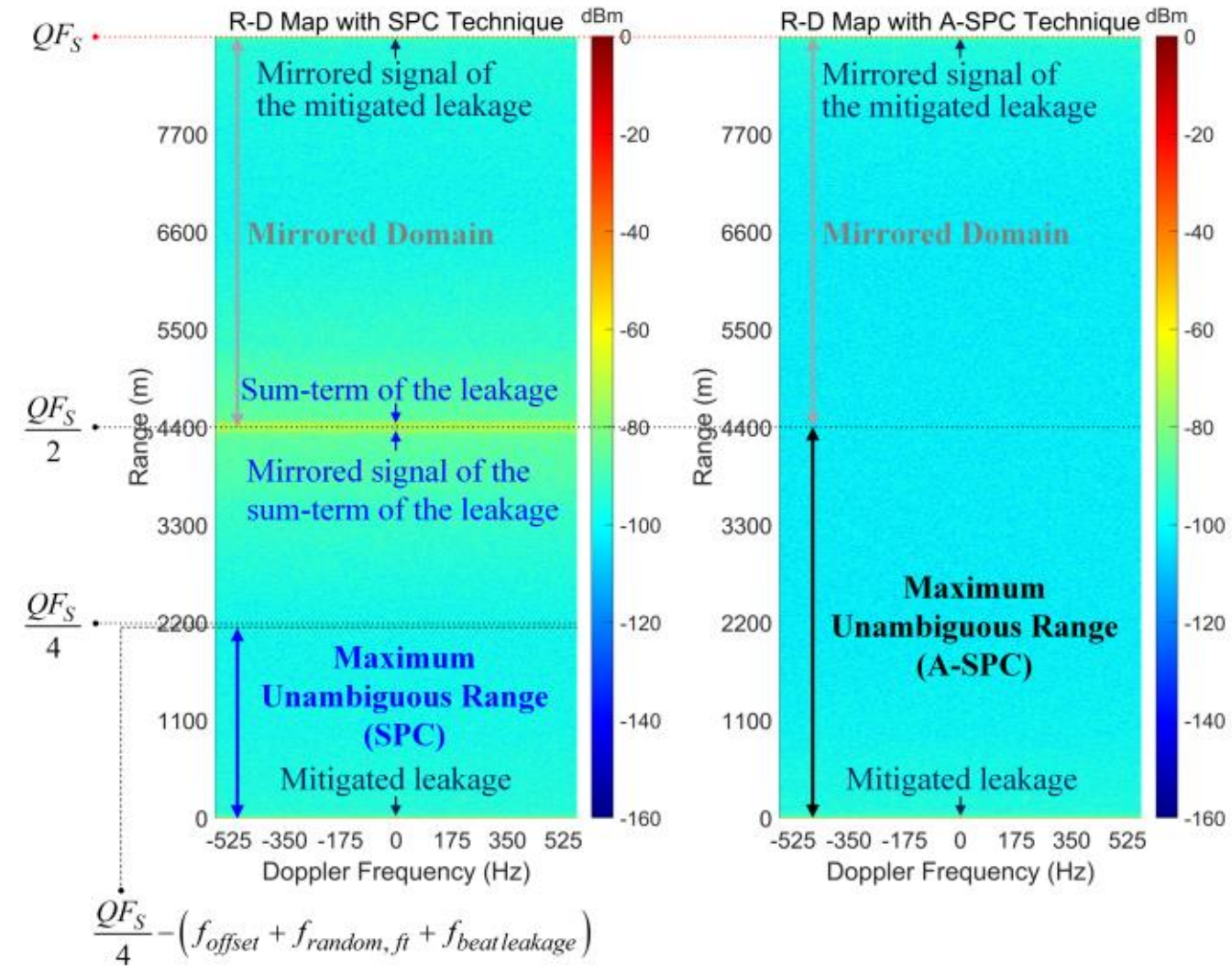

Fig. 15. Results of *Experiment B*: R-D maps. The entire domain of the r-D maps is shown.

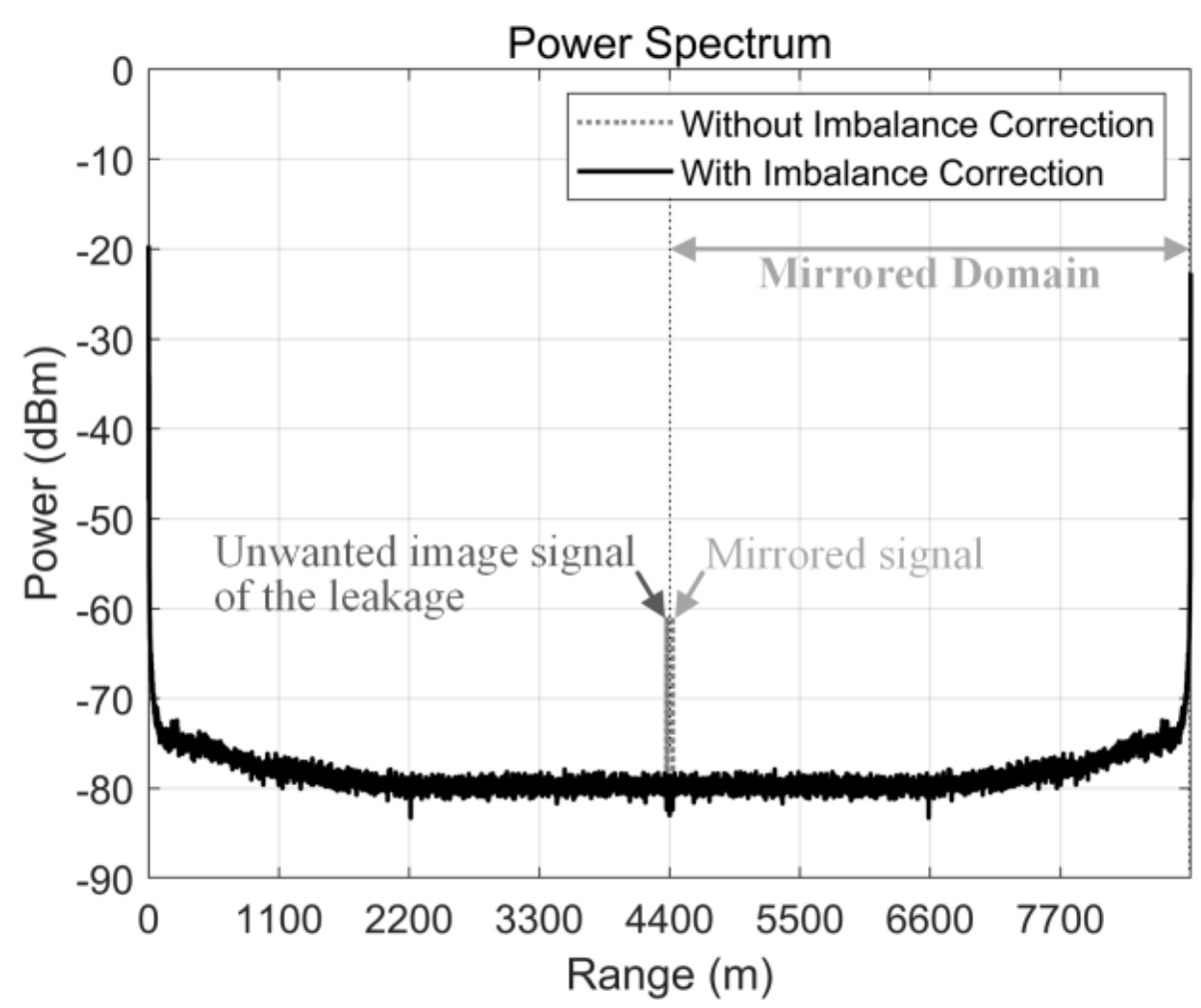

Fig. 16. Results of *Experiment B*: Performance of the quadrature imbalance correction.

## B. *Results and Discussion for Experiment B*

The results of *Experiment B* are shown in Fig. 14-17. Fig. 14 and Fig. 15 show the resulting power spectra and r-D maps. In the case of the SPC technique, the maximum unambiguous range (MUR), which is free from aliasing, is less than a quarter of the entire domain due to the sum-term. Besides, the larger the values of $f_{offset}$, $f_{random,ft}$, and $f_{beat\ leakage}$, the smaller the MUR in the case of the SPC technique. On the other hand, because the A-SPC technique does not produce any sum-terms, the MUR becomes the half of the entire domain. Also, there are no concerns that the MUR will decrease due to $f_{offset}$, $f_{random,ft}$ , and $f_{beat\ leakage}$ . Therefore, if the same oversampling frequency is used, the MUR in the A-SPC technique is more than twice the MUR in the SPC technique.

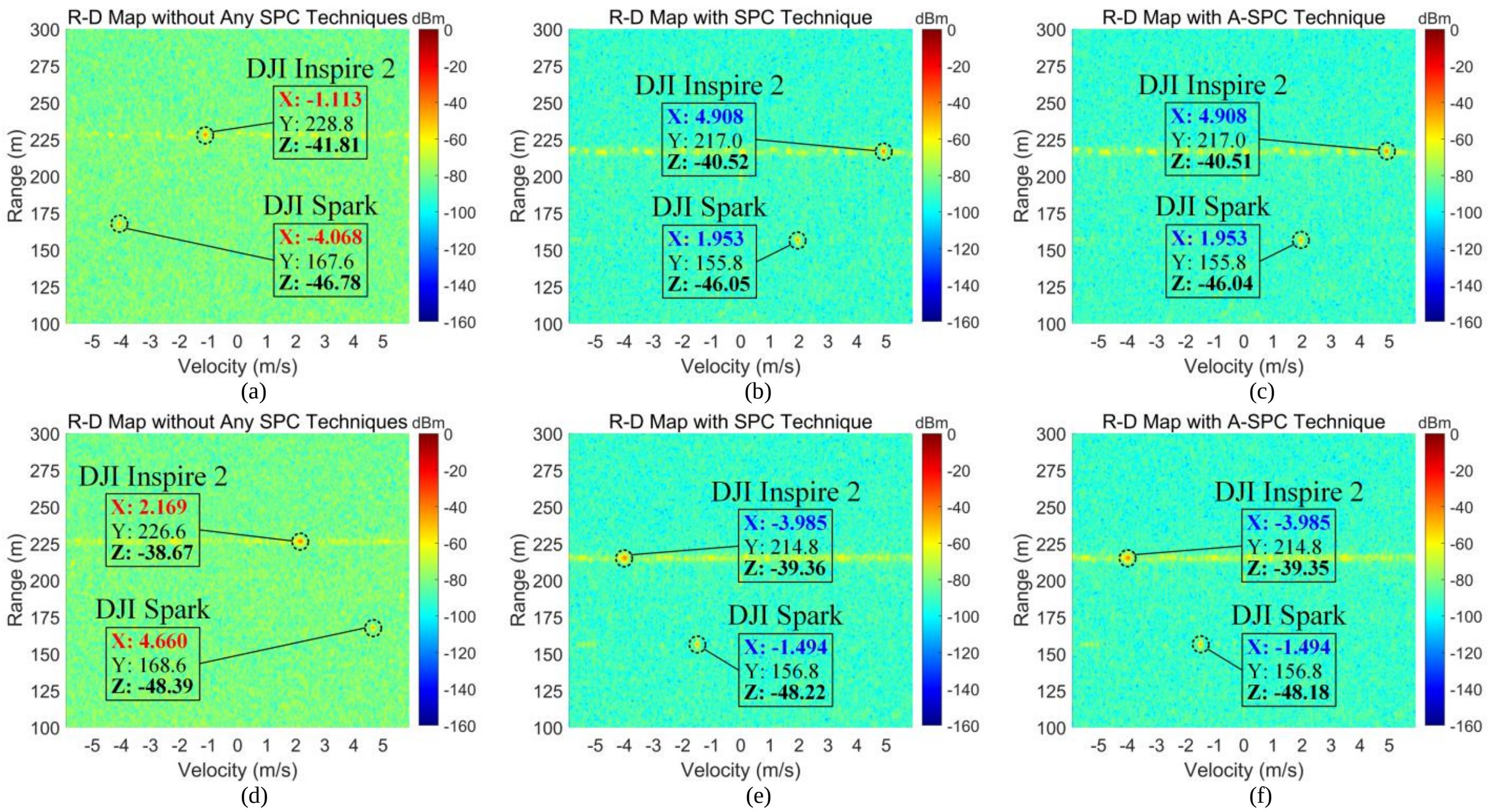

Fig. 17. Results of *Experiment B*: R-D maps when the small moving drones are detected. (a)-(c) R-D maps when the small drones are ascending. (d)-(f) R-D maps when the small drones are descending.

Fig. 16 shows the performance of the quadrature imbalance correction in the A-SPC technique. When the quadrature imbalance is not included in the A-SPC technique, the unwanted image signal of the leakage, which has not negligible power, occurs. On the other hand, the unwanted image signal does not appear in the power spectrum when the quadrature imbalance correction is included in the A-SPC technique. To evaluate the quadrature imbalance correction method, we measured the image rejection ratio (IRR) that can be calculated as follows:

$$IRR = \frac{1 + A_E^2 + 2A_E \cos(\theta_E)}{1 + A_E^2 - 2A_E \cos(\theta_E)}. \tag{13}$$

The averaged *IRR* without the quadrature imbalance correction was about 52.62 dB, while the averaged *IRR* with the quadrature imbalance correction was about 89.82 dB. Therefore, the IRR was improved about 37.2 dB due to our quadrature imbalance correction method.

Fig. 17 shows the resulting r-D maps when the small moving drones are detected. In the r-D maps that any SPC techniques were not applied, which are in Fig. 17(a) and Fig. 17(d), wrong velocity information are measured. On the other hand, not only the SPC technique, Fig. 17(b) and Fig. 17(e), but also the A-SPC technique, Fig. 17(c) and Fig. 17(f), corrects the velocity error and provides accurate velocity information of the small moving drones. Also, the background color of the r-D maps that is the 2-D noise floor level becomes bluer through both the SPC technique and the A-SPC technique, while the power values of the small moving drones remains almost the same. Therefore, it has been verified that the A-SPC technique can also improve the r-D map by increasing the 2-D SNR and the accuracy of the measured velocity of the small moving drones.

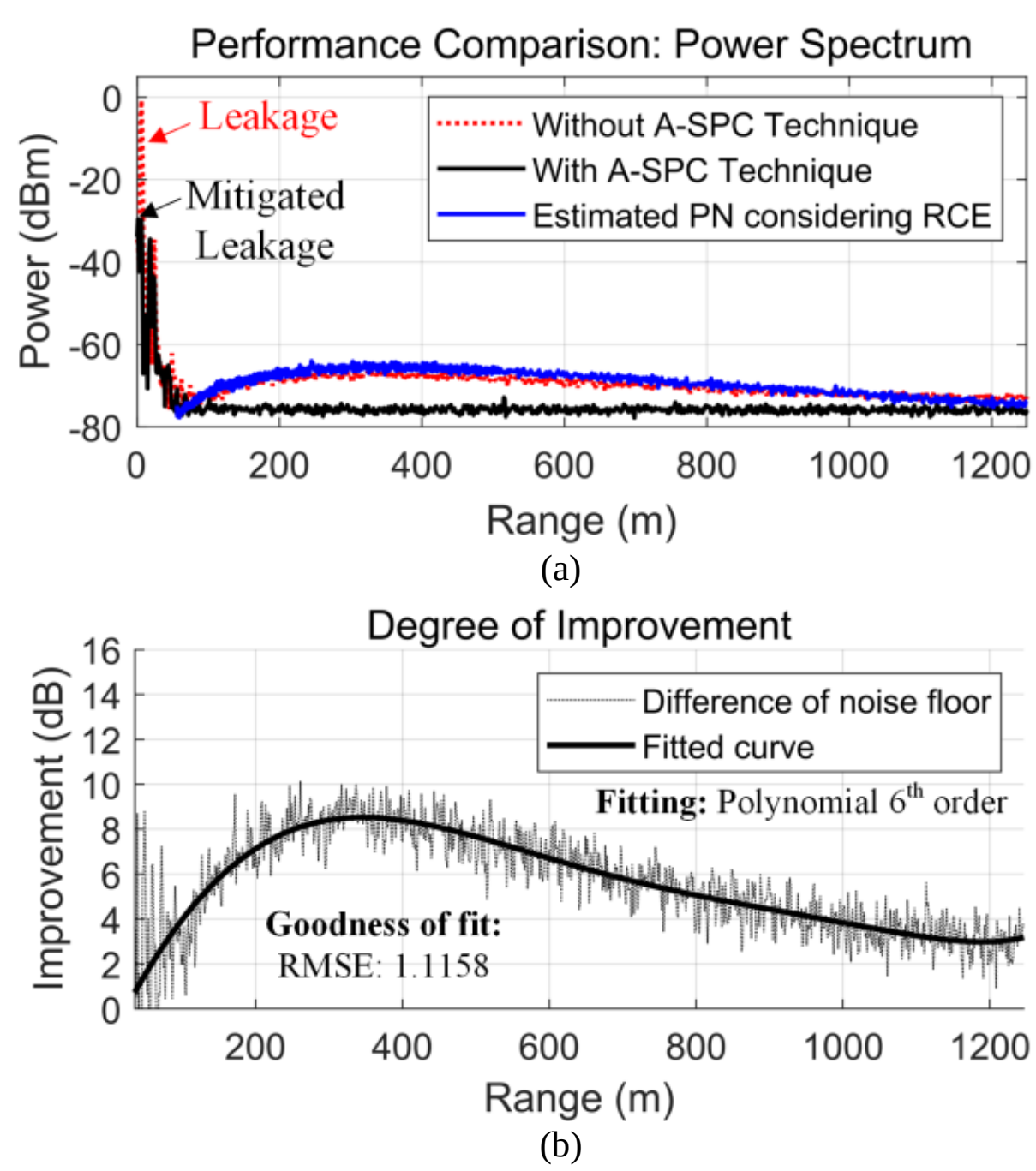

Fig. 18. Results of *Experiment C*: Power spectra. (a) Performance comparison of the power spectra. (b) Degree of improvement.

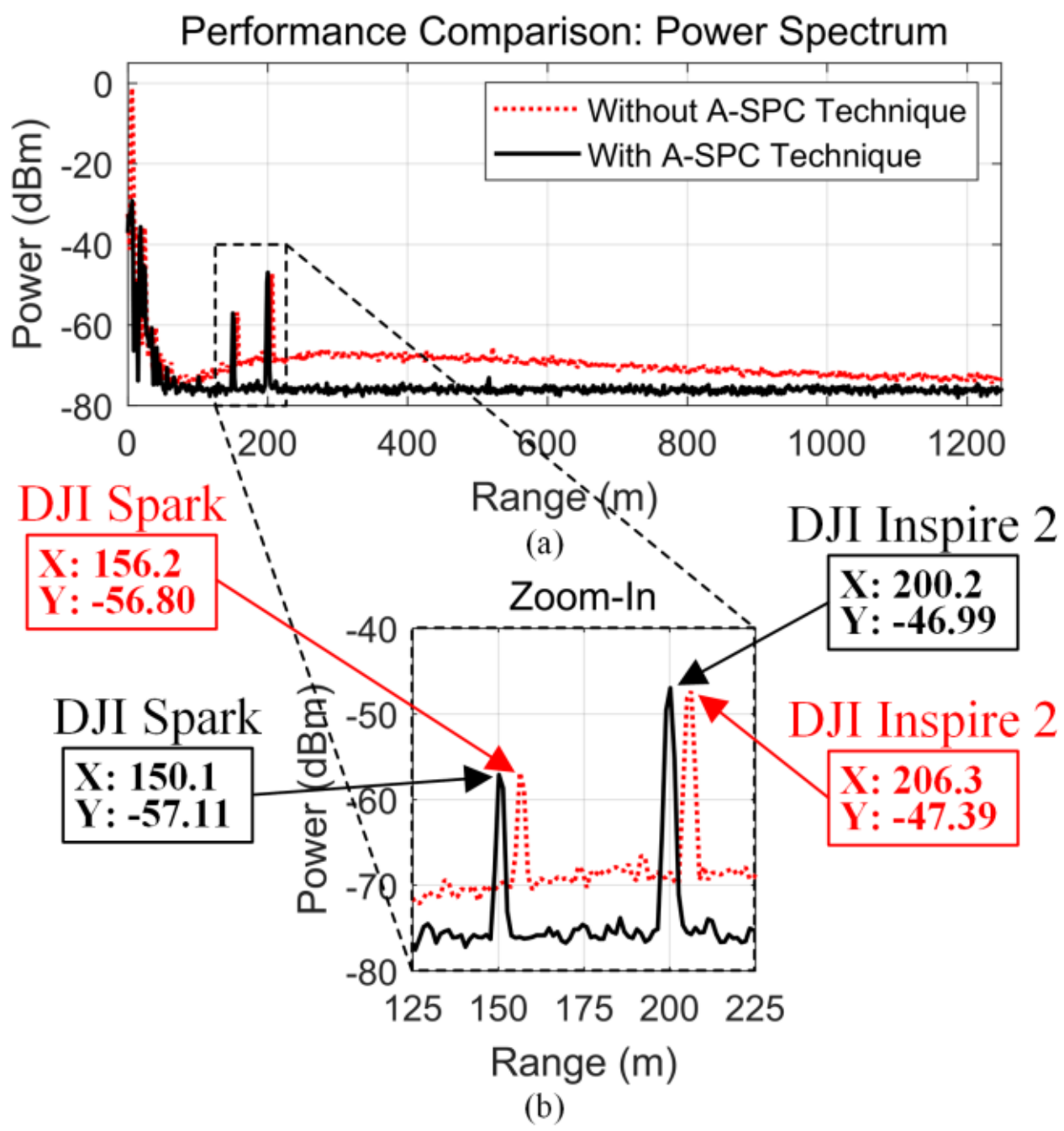

Fig. 19. Results of *Experiment C*: Power spectra when the small hovering drones are detected. (a) Performance comparison of the power spectra. (b) Zoomed in version of (a).

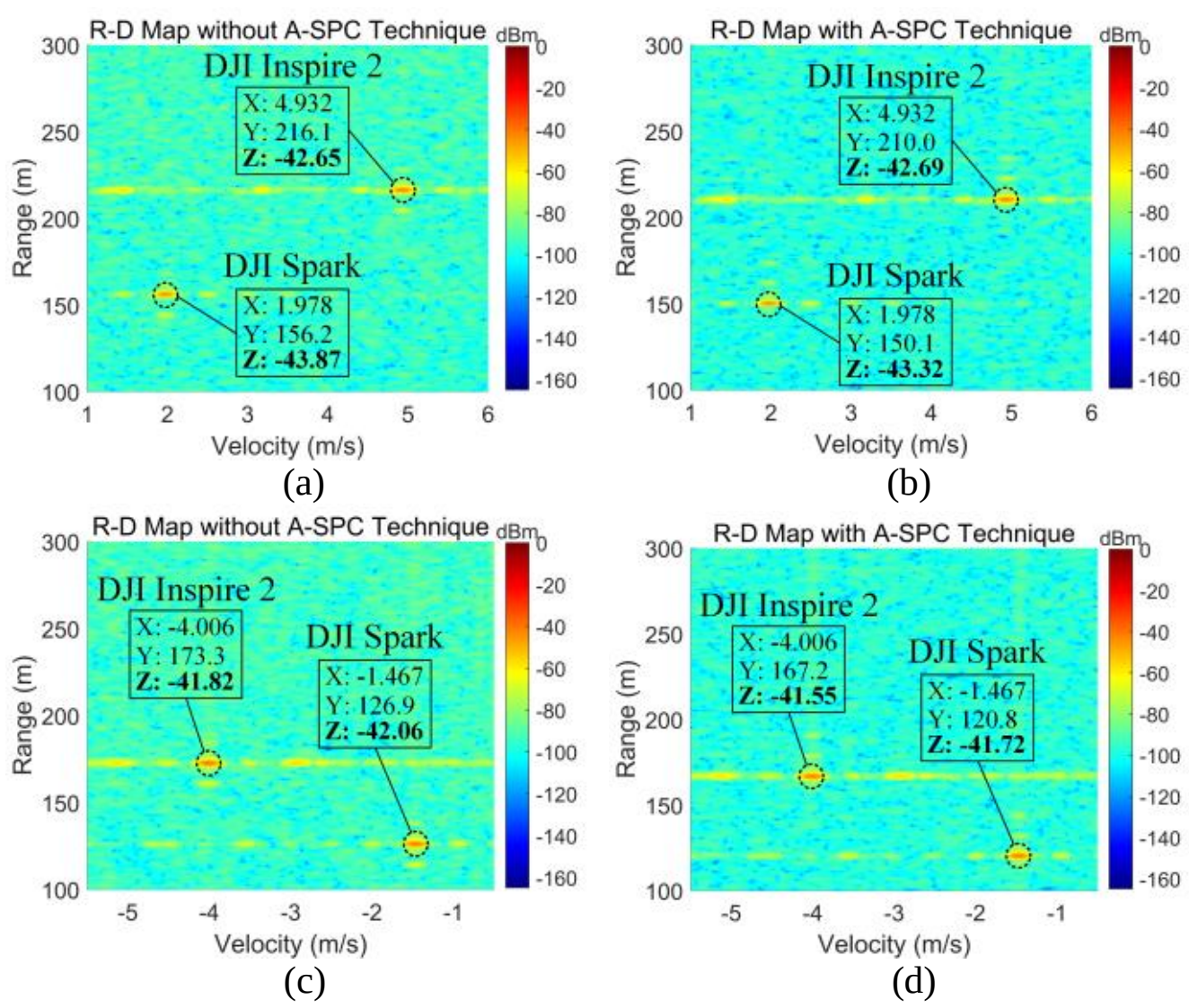

Fig. 20. Results of *Experiment C*: R-D maps when the small moving drones are detected. (a)-(b) R-D maps when the small drones are ascending. (c)-(d) R-D maps when the small drones are descending.

### C. *Results and Discussion for Experiment C*

The results of *Experiment C* are shown in Fig. 18-20. The Fig. 18(a) shows the results of power spectra. The estimated phase noise based on the measured phase noise and considering the RCE is well matched with the noise floor. Therefore, it was confirmed that the RCE effect is applied and the phase noise of the leakage with the RCE effect dominates the noise floor. The degree of improvement is shown in Fig. 18(b). Although the shape of the phase noise is changed, the A-SPC technique performs the effect of the leakage mitigation by attenuating the phase noise of the leakage regardless of the change in its shape.

TABLE IV
COMPARISON OF SPC TECHNIQUE AND A-SPC TECHNIQUE

| Features | [6]-[9] | This work |
|---|---|---|
| Technique | SPC technique | A-SPC technique |
| Strategic frequency planning | Required | Not required |
| Oversampling | Required | Not required |
| Noise floor and SNR improvement | Good | Better |
| Maximum unambiguous range ($R_{MUR}$) | $2R_{MUR,SPC} < R_{MUR,A-SPC}$ (when the same oversampling frequency is applied) | |
| Range error correction (Internal Delay Compensation) | Possible | Possible |
| Doppler error correction (Phase Calibration) | Possible | Possible |
| Available radar architecture | Limited | Not limited |

Fig. 19 shows the power spectra when the small hovering drones are detected. The A-SPC technique improves the SNR by reducing the noise floor and maintaining the signal powers of the small drones. Moreover, referring the true distance values in Fig. 9(b), the A-SPC technique provides the distance information of the small drones with high accuracy, while the distance information are not correct in the power spectrum without the A-SPC technique.

The r-d maps when the small moving drones are detected are shown in Fig. 20. As we have observed in *Experiment B*, while the signal powers of the small moving drones remain almost the same, the background color is bluer in the r-D maps with the A-SPC technique, Fig. 20(b) and Fig. 20(d), than in the r-D maps without the A-SPC technique, Fig. 20(a) and Fig. 20(c). Thus, it has been proved that the A-SPC technique can also increase the 2-D SNR of the small moving drones. Note that unlike the heterodyne FMCW radar, $f_{offset}$ is not generated in the homodyne FMCW radar due to its direct down-conversion. Therefore, there are no unwanted Doppler shift in homodyne FMCW radars, if $f_{random,st}$ does not exist like our homodyne FMCW radar. Even if there are homodyne FMCW radars that have $f_{random,st}$ and the unwanted Doppler shift, it is obvious that the A-SPC technique can correct the velocity errors due to the unwanted Doppler shift, because the A-SPC technique has verified this effect in the heterodyne FMCW radar through *Experiment B*.

Through *Experiment C*, we have demonstrated that the A-SPC technique works well even in the homodyne FMCW radar and performs its functions well.

## V. CONCLUSION

In this paper, we have proposed the A-SPC technique to overcome the limitations of the SPC technique and verified its performances. The comparisons between the SPC technique and the proposed A-SPC technique are summarized in Table IV. The SPC technique requires the strategic frequency planning and the oversampling and cannot be applied in the homodyne FMCW radar. Unlike the SPC technique, the A-SPC technique can be realized without the strategic frequency planning and the oversampling, thus it does not limit the freedom in radar system design and does not require the high-performance ADC and memories. Also, the A-SPC technique can even increase the degree of improvement in the noise floor. In addition, of course, the oversampling is not required in the A-SPC technique, but if the same oversampling frequency is used, the MUR resulting from the A-SPC technique is more than twice the MUR resulting from the SPC technique. The r-D map can be extracted through the A-SPC technique, and the effects of range and Doppler error correction in the SPC technique are maintained even in the A-SPC technique, thus the A-SPC technique can also provide the accurate distance and velocity information of the targets. Finally, we have demonstrated that the A-SPC technique can be applied even in the homodyne FMCW radar, unlike the SPC technique. Therefore, the A-SPC technique is more efficient and powerful than the SPC technique, so it is highly useful in practice.